# sDTW: Computing DTW Distances using Locally Relevant Constraints based on Salient Feature Alignments*

K. Selçuk Candan
Arizona State University
Tempe, AZ, USA
candan@asu.edu

Rosaria Rossini
University of Torino
Torino, Italy
rossini@di.unito.it

Maria Luisa Sapino
University of Torino
Torino, Italy
mlsapino@di.unito.it

Xiaolan Wang
Arizona State University
Tempe, AZ, USA
xwang220@asu.edu

## ABSTRACT

Many applications generate and consume temporal data and retrieval of time series is a key processing step in many application domains. Dynamic time warping (DTW) distance between time series of size $N$ and $M$ is computed relying on a dynamic programming approach which creates and fills an $N \times M$ grid to search for an optimal *warp path*. Since this can be costly, various heuristics have been proposed to cut away the potentially unproductive portions of the DTW grid. In this paper, we argue that time series often carry structural features that can be used for identifying *locally relevant* constraints to eliminate redundant work. Relying on this observation, we propose *salient feature* based sDTW algorithms which first identify robust salient features in the given time series and then find a consistent alignment of these to establish the boundaries for the warp path search. More specifically, we propose alternative *fixed core&adaptive width*, *adaptive core&fixed width*, and *adaptive core&adaptive width* strategies which enforce different constraints reflecting the high level structural characteristics of the series in the data set. Experiment results show that the proposed sDTW algorithms help achieve much higher accuracy in DTW computation and time series retrieval than *fixed core & fixed width* algorithms that do not leverage local features of the given time series.

## 1. INTRODUCTION

Since many applications generate and/or consume temporal data, querying and clustering of sequences and time series have been core data operations in many application domains, from speech recognition, intrusion detection, to finance (Figure 1). As a consequence, there has been significant amount of research both into defining measures for comparing sequences and sub-sequences, as well as into the development of efficient data structures and algorithms for implementing these core operations [1,3,6,16].

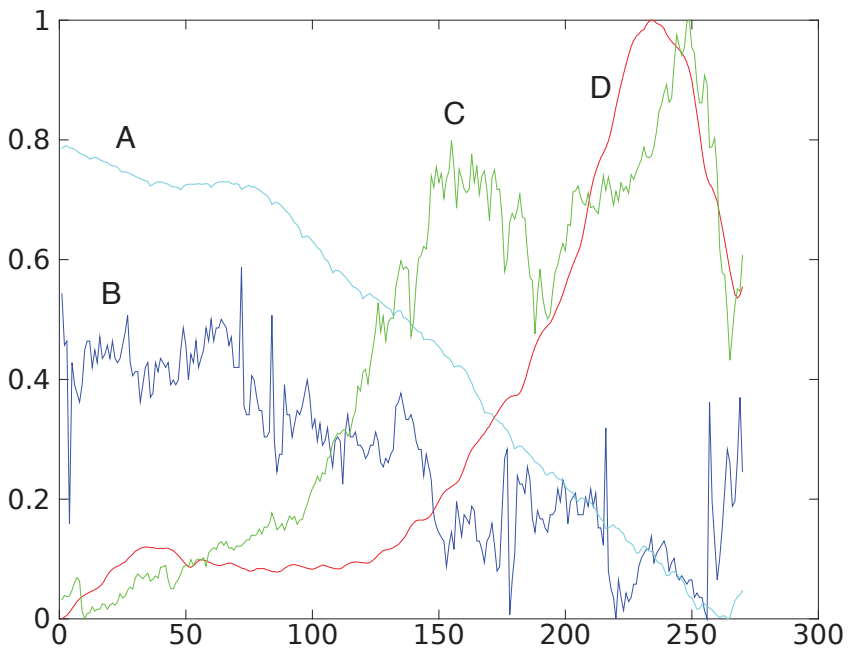

**Figure 1: Four sample economic index time series (obtained from [14]): note that series A and B are similar to each other and different from the others (similarly for the pair C and D)**

In most applications, when comparing two sequences or time series, exact alignment is not required (Figure 1). Instead, whether two sequences are going to be treated as matching or not depends on the amount of difference between them; thus, this difference needs to be quantified. This is commonly done through *distance* measures which quantify the minimum number (or cost) of symbol *insertions*, *deletions*, and *substitutions* needed to convert one sequence to the other [4,10]. *Dynamic time warping (DTW)* distance [2,3,7,8,16], used commonly when comparing *continuous sequences* or time series (especially in scenarios where the series carry similar underlying patterns, but are different from each other due to temporal deformations, such as shifts and stretches), can be thought of as a special case of this general edit distance measure.

As shown in Figure 2(a), and described later in Section 2.1, given two time series $X$ (of length $N$) and $Y$ (of length $M$), the DTW distance is often defined as the length of the shortest *warp path* that can be used for aligning $X$ and $Y$. Intuitively, the warp path (which can be visualized as a path from the lower-left corner of an $N \times M$ grid to its upper-right corner as shown in Figure 2(a)) describes for each element of $X$, the corresponding continuous stretch of $Y$ (consisting of one or more elements), and vice versa. The search for an optimal warp path on the grid is commonly performed using an $O(NM)$ dynamic programming based algorithm, which first fills the cells in the grid from the lower-left corner to the upper-right corner with partial DTW path costs and then, once the grid is filled, selects that path by traversing back to the lower-left corner in $O(N + M)$ time. Since $O(NM)$ execution cost is often too high, various heuristics which prune the grid have been proposed (Figure 2(b,c)). These heuristics impose various constraints on the positions which can be explored during the filling of the grid.

*This work is supported by the NSF Grant 1043583 " MiNC: NSDL Middleware for Network- and Context-aware Recommendations". Authors are listed in alphabetical order.

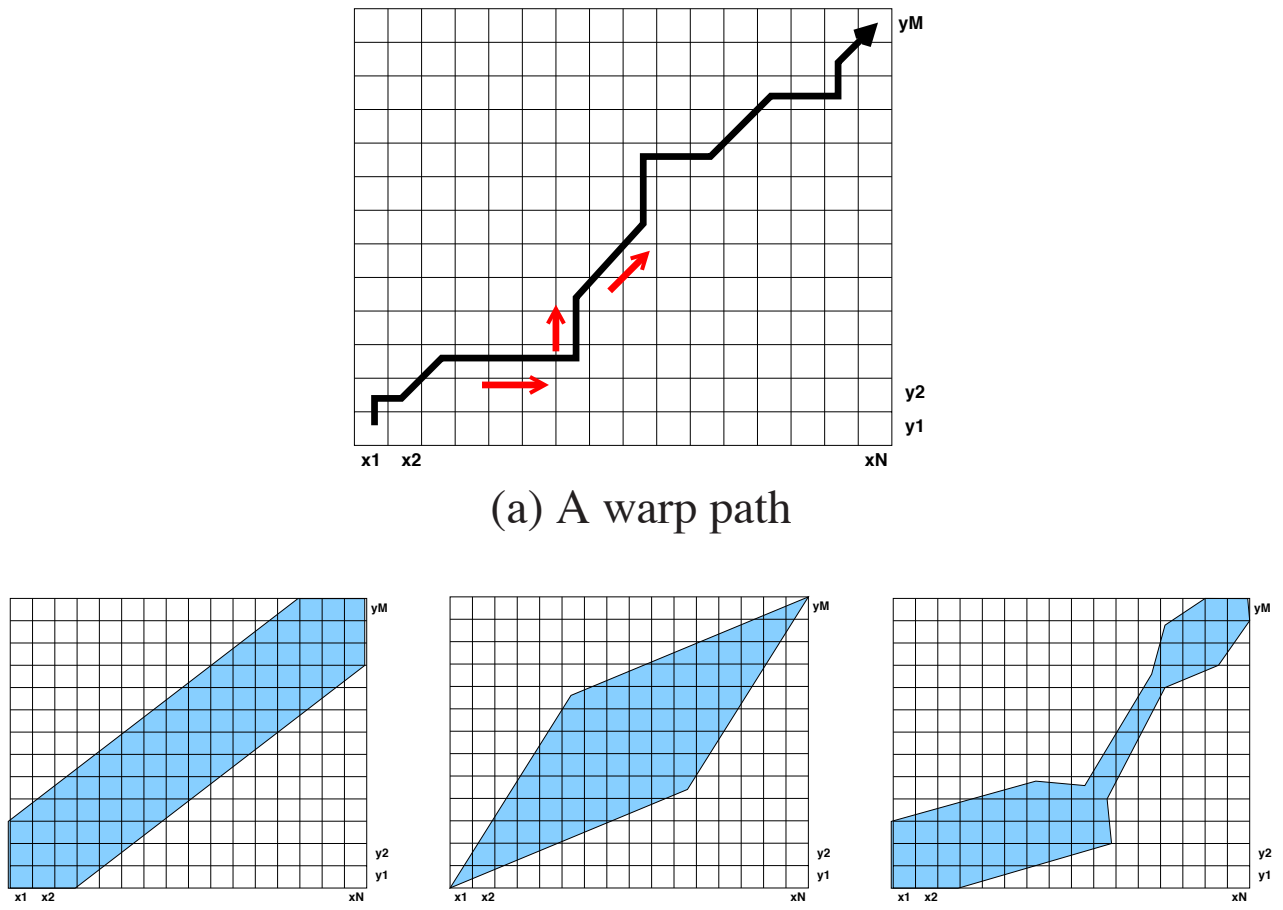

(a) A warp path

(b) Sakoe-Chiba (c) Itakura paral. (d) sDTW band

**Figure 2: (a) A warp path on the $N \times M$ grid; and the shapes of the bands imposed by (b) Sakoe-Chiba, (c) Itakura parallelogram, and (d) sDTW constraints: shapes of the sDTW search bands are adapted using locally relevant constraints discovered using salient-features of the input series**

A key difficulty in constraining the DTW grid, however, is that the tighter the constraints are the more likely the optimal warp path will be missed. It is easy to see that a tighter constraint (e.g., a thinner Sakoe-Chiba band in Figure 2(b)) would speed up the DTW computation, but would also reduce accuracy; therefore, the constraints must be carefully selected. [15], for example, presents an adaptive technique which learns the appropriate sizes of the band at different portions of the input time series and thus attempts to strike a balance between speed and quality. This approach to informing the DTW pruning process, however, requires training data (in the form of user feedback on data samples) and a learning process that can determine the relevant constraints at different parts of the DTW grid. In contrast, the *salient feature* based sDTW algorithms presented in this paper rely on alignment evidences provided by *salient features* in the time series themselves to improve the effectiveness of the pruning constraints.

As discussed later in Section 2, an orthogonal approach to reducing work involves relying on reduced representations of the data and iteratively improving results [2,8,18]. sDTW algorithms operate in the original time resolution, but nevertheless leverages temporal features of multiple scales. Moreover, it can naturally be combined with reduced representation based solutions.

## 1.1 Contributions of this Paper: sDTW – Locally Relevant DTW Pruning Constraints based on Salient Feature Alignments

In this paper, we recognize that in many cases the two time series that are being compared carry sufficient structural evidences (in the form of *salient temporal features*) that can be used for helping set *locally relevant* constraints to guide the search for the optimal warp path. Relying on this observation, we propose *salient feature* based sDTW algorithms that first locate *salient features* on the time series that are robust against various types of noise and then use these salient features as evidences to extract *locally relevant* constraints that adapt the shape of the DTW search band (Figure 2(d)). The structure of the paper and our key contributions are as follows:

1. In Section 2, we introduce the relevant background, including the key concepts used in the paper and an overview of the related work.
2. In Section 3.1, we describe a method for locating *salient features* on the time series that is robust against various types of noise. The proposed method relies on a *scale-invariant feature transform* based feature extraction process (as in SIFT [11,12]) for locating robust salient feature points on time series and, then, extracts temporal feature descriptors that can be used for searching for salient feature alignments across time series. However, unlike the basic SIFT algorithms, which are optimized for extracting rough object features from 2D images, the salient feature search algorithms and salient feature descriptors we present are optimized for supporting precise alignments of 1D time series. These salient features not only describe the temporal position of the located salient features, but also their temporal scales.
3. In Section 3.2, we describe how to find consistent *salient alignments* of a given pair of time series by first matching the salient features and then eliminating inconsistencies in a way that maximizes the quality of structural alignments.
4. Finally, in Section 3.3, we describe how to use the discovered robust salient alignments to compute locally relevant constraints that can effectively prune the DTW grid (Figure 2(d)). In particular, we present three different sDTW constraint types (*fixed core&adaptive width*, *adaptive core&fixed width*, and *adaptive core&adaptive width*) which use the available alignment evidences to inform the search for the DTW path in different ways.

In Section 4, we experimentally evaluate the impact of the various salient feature-based locally relevant pruning constraints described in this paper on the efficiency and effectiveness of retrieval and classification of time series.

# 2. BACKGROUND AND RELATED WORK

In this section, we present the background, including the key concepts used in the paper, and an overview of the related work in the literature.

## 2.1 Dynamic Time Warping (DTW)

Dynamic time warping (DTW) is a common technique for comparing sequences or time series by searching for optimal alignments, described in terms of, so called, *warp paths*.

### 2.1.1 Warp Path

Let us be given two sequences or time series, $X = (x_1, x_2, \ldots, x_N)$ and $Y = (y_1, y_2, \ldots, y_M)$, where $x_i$ and $y_j$ are from the same domain $D$, and let $\Delta()$ be a distance function for comparing elements in $D$. An alignment from $X$ to $Y$ is described in terms of a *warp path* $W = (w_1, w_2, \ldots, w_K)$, where

- $max(N, M) \leq K \leq N + M$,
- $w_1 = (1, 1)$,
- $w_K = (N, M)$, and
- $w_l - w_{l-1} \in \{(1,0), (0,1), (1,1)\}$.

The warp path $W$ between $X$ and $Y$ can be visualized as a path from the lower-left corner of an $N \times M$ grid to its upper-right corner, where the path is constrained to *monotonically advance* in horizontal or vertical direction (or both) in each step (Figure 2(a)).

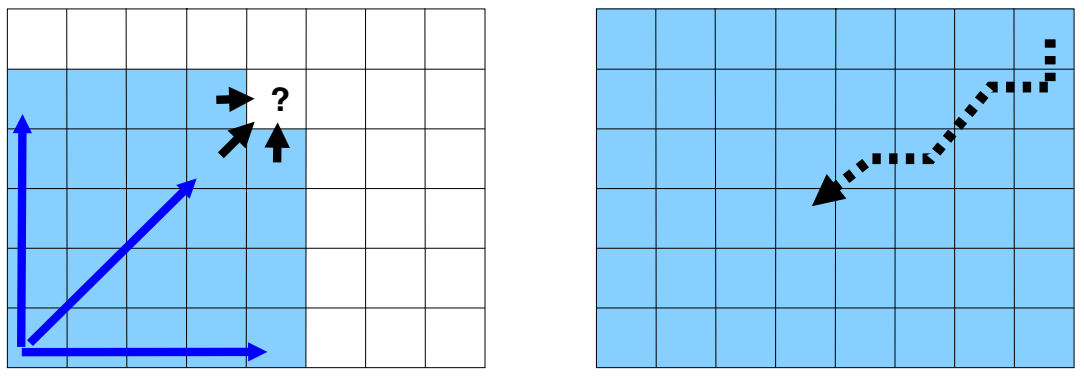

(a) Filling of the DTW grid (b) Search for the warp path

**Figure 3: (a) Visualization of (a) the dynamic programming algorithm to fill the $D$ matrix and (b) the algorithm for identifying the optimal warp path**

#### 2.1.2 DTW Distance

The *overall distance* of a given warp path, $W = (w_1, w_2, \ldots, w_K)$, between time series $X$ and $Y$ is defined as

$$\Delta(W) = \sum_{l=1}^{K} \Delta(x_{w_l[1]}, y_{w_l[2]}).$$

Given this, an *optimal alignment* is defined as a warp path over the time series $X$ and $Y$ with the minimum overall distance over all possible warp paths. The goal of DTW algorithm is to find this *optimal* alignment between $X$ and $Y$; in other words, the DTW distance between $X$ and $Y$ is defined as

$$\Delta_{DTW}(X,Y) = min\{\Delta(W)\ |W is\ a\ warp\ path\ for\ X, Y\}.$$

Note that the DTW distance is symmetric, but does not necessarily satisfy the triangular inequality – thus, it is not a metric.

#### 2.1.3 Dynamic Programming based Computation of the DTW Distance

Since testing all possible warp paths for $X$ and $Y$ would be prohibitively expensive, like many other edit distance measures, the DTW distance is also commonly computed by leveraging the underlying recursive nature of the distance function: Let $X(1:i)$ denote the $i$-length prefix of $X$, $Y(1:j)$ denote the $j$-length prefix of $Y$, and $D(i,j)$ be defined as $\Delta_{DTW}(X(1:i), Y(1:j))$. Then, it can be shown [7], that $D(i,j)$ is equal to

$$min\{D(i-1,j), D(i,j-1), D(i-1,j-1)\} + \Delta(x_i, y_j).$$

Consequently, the value of $\Delta_{DTW}(X,Y) = D(N,M)$ and the corresponding optimal warping path $W^{opt}$ can be identified using a dynamic programming algorithm that fills the $(N+1)\times(M+1)$ matrix, $D$, in a bottom up fashion starting from $D(0,0)$ in $O(NM)$ time (Figure 3).

#### 2.1.4 Speeding Up the DTW Computation

While being significantly faster than an exhaustive enumeration of all possible warp paths between $X$ and $Y$, the $O(NM)$ time needed to fill the underlying accumulation matrix, $D$, is still unacceptably high for many applications. Therefore, various optimization and approximation algorithms have been proposed [5,8,15,17]. Most of these algorithms speed the process by pruning the accumulation matrix $D$, thereby reducing the number of operations needed to fill it during the dynamic programming process. The *Sakoe-Chiba band* approach [17], for example, constrains the feasible regions through which the warp path can pass to a (relatively) narrow path along the diagonal (Figure 2(b)). The *Itakura parallelogram* [5], in contrast, places a constraint on the slope of the path (Figure 2(c)).

An orthogonal approach to speeding up the DTW computations is to rely on a reduced representation, where one first identifies a warp path at a low resolution data and then improves on this by further refining the warping path at higher resolutions [2, 8, 18]. In this paper, we present a constraint based solution, which nevertheless is able to leverage temporal features at multiple scales of time. Note that the proposed approach can naturally be implemented along with reduced representation based solutions.

## 3. SDTW: SPEEDING UP DTW COMPUTATION THROUGH SALIENT FEATURE ALIGNMENTS

As described in Section 1.1, our goal in this paper is to leverage *salient alignment evidences* to improve the effectiveness of the pruning constraints. In particular,

1. we first search for *salient features* on the input time series that are robust against various types of noise,
2. then, we find *salient alignments* of a given pair of time series by matching the descriptors of the salient features, and
3. finally, we use these salient alignments to compute locally relevant constraints (on the band size and slope) to prune the warp path search (Figure 2(d)).

In this section, we describe each of these three steps in detail.

### 3.1 Searching for Salient Features

In many applications, not the global features of the whole data objects, but the local features of the salient parts of the objects are more relevant for effective retrieval or classification.

#### 3.1.1 Scale-Invariant Feature Transform (SIFT)

The scale-invariant feature transform (SIFT) [11, 12] algorithm relies on this observation for salient point (or *keypoint*) based search of 2D images: the algorithm identifies keypoints of a given image (and their descriptors) that are invariant to image scaling, translation, rotation, and also partially invariant to illumination differences and noise. It has been shown that SIFT-based local descriptors perform very well in the context of matching and recognition of the same scene or object under different viewing conditions [13]. The algorithm relies on a four step process to identify such salient points and their robust local feature descriptors in 2D images:

**Step 1: Scale-space extrema detection:** The first stage of the process identifies candidate points of interest, $\langle x, y, \sigma\rangle$ across multiple scales of the given image (here $\sigma$ denotes the image scale) by searching over multiple scales and locations of the given image. These candidate points of interest are those points with the largest variations with respect to their neighbors in both space and scale. Let $D(x, y, \sigma)$ be the difference between the versions of the input image smoothed at different scales, $\sigma$ and $k\sigma$ (for some constant multiplicative factor $k$). To detect the local maxima of $D(x, y, \sigma)$, each pixel is compared with its neighbors at the same scale as well as neighbors at images up and down one scale.

**Step 2: Keypoint filtering and localization:** At the next step, those candidate points identified in the first step that are sensitive to noise are eliminated. These include those points that have low contrast or are poorly localized along edges in the image.

**Step 3: Orientation assignment:** At the third step, a dominant orientation, $o$, is assigned to each remaining keypoint, $\langle x, y, \sigma\rangle$, based on the local histogram distribution.

**Step 4: Keypoint descriptor creation:** In the final step of SIFT, for each keypoint, a local image descriptor that is invariant to both illumination and viewpoint is extracted using the location and orientation information obtained in the previous steps. The algorithm samples image gradient magnitudes and orientations (relative to the

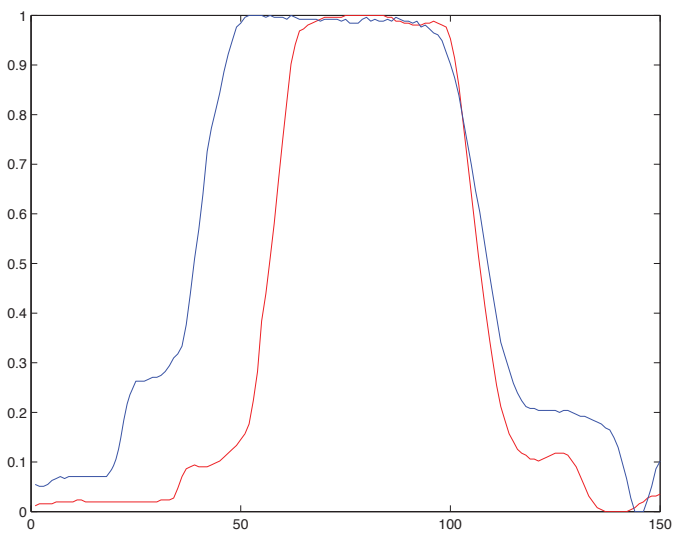

(a) Two time series

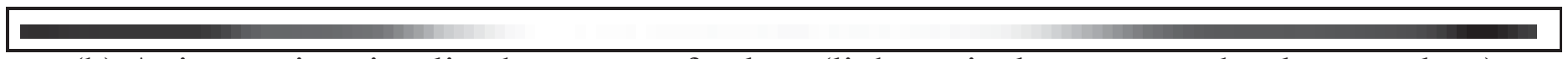

(b) A time series visualized as a row of values (lighter pixels correspond to larger values)

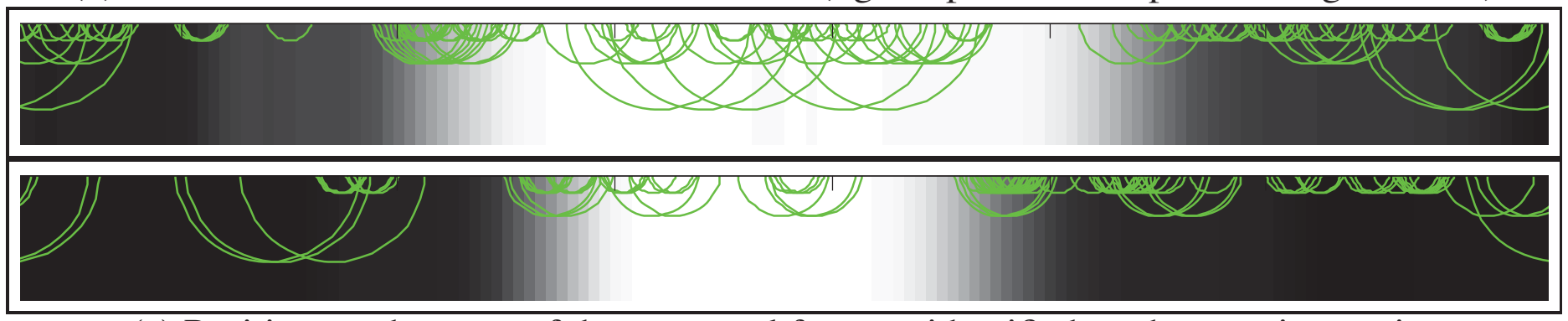

(c) Positions and scopes of the structural features identified on the two time series

**Figure 4: (a) Two time series; (b) one of these series visualized as a single row of values: the white region corresponds to the peak of the time series and the dark regions correspond to the dips on the curve; and (c) the features identified on the two series – each half circle represents a feature and the diameter of the circle represents the scope of the feature**

orientation $o$ identified in the third step) around the keypoint location, $\langle x, y \rangle$, using the scale, $\sigma$, of the keypoint to select the level of the Gaussian blur of the image. To avoid sudden changes in the descriptor with small changes in the position and to give less emphasis to gradients that are far from the center of the descriptor, a Gaussian weighing function is used to assign a weight to the magnitude of each sample point based on its distance from the keypoint.

In this paper, we note that the local salient features identified by SIFT-like algorithms can also be very effective in computing DTW distances between 1D time series: boundaries and descriptors of salient features of time series can be identified through a scale-invariant analysis and these salient features can be used to inform the DTW process. In fact, a key advantage of the SIFT approach is that it can be used to identify not only the center positions of the temporal features, but also the sizes of the features – greatly improving opportunities for more precise alignment of time series.

### 3.1.2 *Salient Feature Search in Time Series*

In this paper, we propose to use a *SIFT-like* salient point extraction algorithm to identify robust points on a given time series. This approach has a number of advantages:

- The identified salient points are robust against noise and common transformations, such as temporal shifts.
- Scale invariance enables the extracted salient points to be robust against variations in speed. Also, the temporal scale at which a feature is located gives an indication about the size (or scope) of the temporal feature.
- The identified salient points are also robust against variations in the absolute values of the time series.

Each of these properties can also be independently controlled: i.e., one can turn on/off a particular invariance based on the application. For example, if not applicable, one can turn off or place a bound on invariance against variations on absolute values or can place a bound on difference in scales of the matching features.

One way to leverage SIFT for salient feature detection on a time series of length $N$ would be to create a 2D matrix of size $N \times K$ by replicating the time series as rows of an image and use the basic SIFT algorithm for detecting keypoints on the resulting image. This obviously would be more expensive than identifying salient features directly on the 1D vector corresponding to the given time series (Figure 4(b)) points and creating the corresponding feature descriptors using this 1D vector. Thus, we modify the 2D SIFT algorithm [12] such that features are sought and localized in a scale-invariant manner along only one dimension:

**Step 1: Scale-space extrema detection:** In this step, we search for points of interest, $\langle x, \sigma \rangle$ across multiple scales of the given time series (here $\sigma$ denotes the time scale) by searching over multiple scales and locations of the given series. Let $L(i, \sigma)$, of a given series $X$, be a version of the series smoothed through convolution with the Gaussian,

$G(x, \sigma) = (1/2\pi\sigma^2)e^{-(x^2)/2\sigma^2}$:

$$L(i, \sigma) = G(i, \sigma) * X(i) = G(i, \sigma) * x_i.$$

Intuitively, the Gaussian smoothing can be seen as a multi-scale representation of the given series and thus the differences between the Gaussian smoothed series correspond to differences between the same image at different scales. Thus, this step searches for those points that have largest variations with respect to both time and scale. More specifically, stable keypoints, $\langle x, \sigma \rangle$, are detected by identifying the extrema of the difference series,

$$D(i, \sigma) = L(i, \kappa\sigma) - L(i, \sigma).$$

Here, $D(i, \sigma)$ is the difference between the versions of the input series smoothed at different scales, $\sigma$ and $\kappa\sigma$ (for some constant multiplicative factor $\kappa$). As in SIFT [12], the given time series is incrementally reduced into $o$ octaves, each corresponding to the doubling of the smoothing rate. Each of these $o$ octaves of the scale space is further divided into $s$ levels by constructing intermediary smoothed series obtained by repeatedly convolving the series with Gaussians with parameter $\kappa$ (note that $\kappa^s = 2$). Each resulting series are then subtracted from the series in the adjacent temporal scale to obtain the corresponding difference-of-Gaussian series, $D$, which is then used for searching for salient features at that scale level. Once the $s$ series corresponding to an octave are processed, we downsample the series corresponding to the doubling of $\sigma$ by picking every second pixel. The resulting series forms the basis of the next octave.

The 2D SIFT approach [12] searches the local maxima of $D(i, \sigma)$ by comparing each point with its neighbors at the same scale as well as neighbors at time one up and one down in scale and eliminates all the rest. This, however, implies that features that are similar in scale and time may prune each other. While such pruning may be acceptable and desirable in image search applications, since our goal is to use the located features in fine-tuning the search space for the warp path in DTW, we do not necessarily seek to over-prune the keypoints. Thus, instead of requiring local *maxima* of $D(i, \sigma)$, we accept $\langle x, \sigma \rangle$ as a robust keypoint if it is larger than $(1 - \epsilon)\times$ of each of its neighbors, for a small positive $\epsilon$.

Note that, as shown in Figure 4(c), each identified temporal feature has an associated scope, defined by the temporal scale in which it is located. We set the radius of the scope to $3\sigma$ since, under Gaussian smoothing, 3 standard deviations would cover $\sim 99.73\%$ of the original time points that has contributed to the identified keypoint. Intuitively, the larger the scale is, the bigger the scope of the corresponding temporal feature.

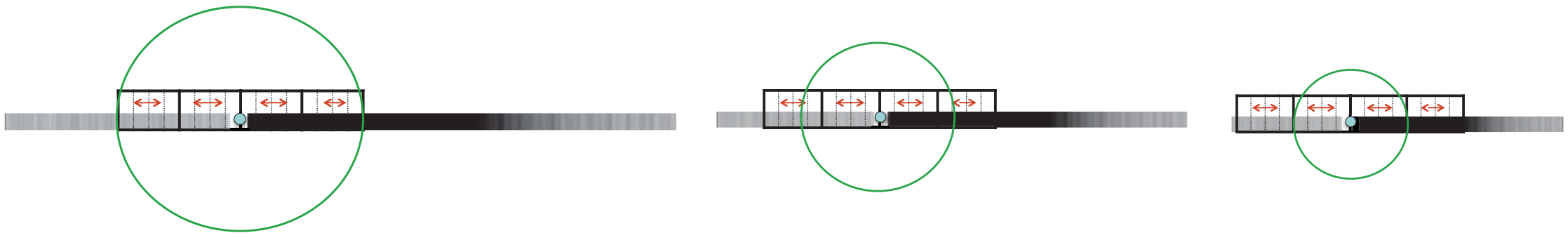

(a) Scale #1 (fine) (b) Scale #2 (medium) (c) Scale #3 (rough)

**Figure 6: The same descriptor size would cover different temporal ranges at different time scales (in this example, the same time series is subjected to three different temporal scales)**

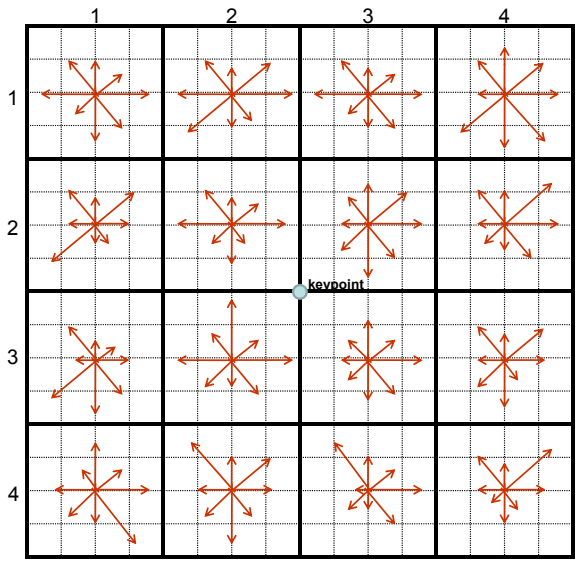

(a) 128 gradients of a keypoint descriptor in SIFT

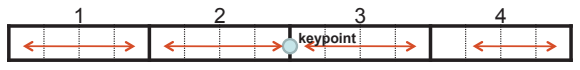

(b) 8 gradients of a salient feature descriptor for a time series

**Figure 5: (a) A 128-dimensional SIFT feature vector describes the gradient distribution around the point $\langle x, y\rangle$ at scale $\sigma$; (b) 8 gradients for a time series describe the magnitude of the changes at different distances from the salient point**

**Step 2: Keypoint descriptor creation:** Next, for each remaining stable point, a descriptor is created by sampling the gradient magnitudes around the salient point, $i$, at time scale, $\sigma$. To avoid sudden changes in the descriptor with small changes in the position of the time window and to give less emphasis to gradients that are far from the center of the descriptor, a Gaussian weighting function is used to assign a weight to the magnitude of each sample point based on its distance from the salient point. The result is a salient feature descriptor describing the local gradients around the salient point on the time series (Figure 5(b)).

In 2D SIFT, each feature descriptor is a vector of length $2a \times 2b \times c$: the descriptor is constructed by superimposing a $2a \times 2b$ grid on top of a 16-pixel by 16-pixel region centered around the keypoint $\langle x, y\rangle$ at scale, $\sigma$. Then, for each pixel in each cell of the grid, the corresponding gradient is computed. Next, for each cell, a $c$-bin gradient histogram is constructed by partitioning the gradient magnitudes in the cell into 8 bins, representing $c$ orientations. For example, in Figure 5(a), we see the gradients represented in a descriptor of length 128 ($= 4 \times 4 \times 8$): in this example, a $4 \times 4$ grid is superimposed on top of the 16-pixel by 16-pixel region around the keypoint and, for each cell of the grid, an 8-bin gradient histogram is constructed, representing 8 major directions. Unlike 2D images, however, in the case of time series the only relevant gradients are along the horizontal direction. Therefore, instead of the $2a \times 2b \times c$ length descriptor as in SIFT, we construct and maintain a $2a \times 2$ length descriptor. This means that, in the example in Figure 5, instead of a descriptor of length 128 ($= 4 \times 4 \times 8$), we would construct a descriptor of length 8 ($= 4 \times 2$).

As shown in Figure 6, the same descriptor size would cover different temporal ranges at different time scales: a given descriptor size covers larger temporal ranges in scales corresponding to reduced representations of time series. Therefore, the descriptor size must be selected in a way that reflects the temporal characteristics

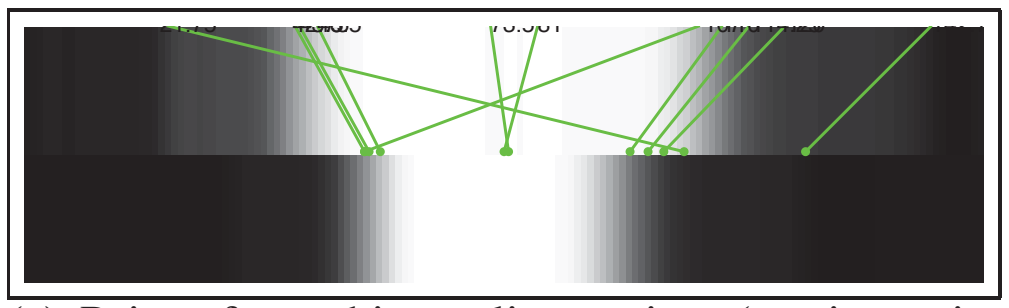

(a) Pairs of matching salient points (pre-inconsistency removal)

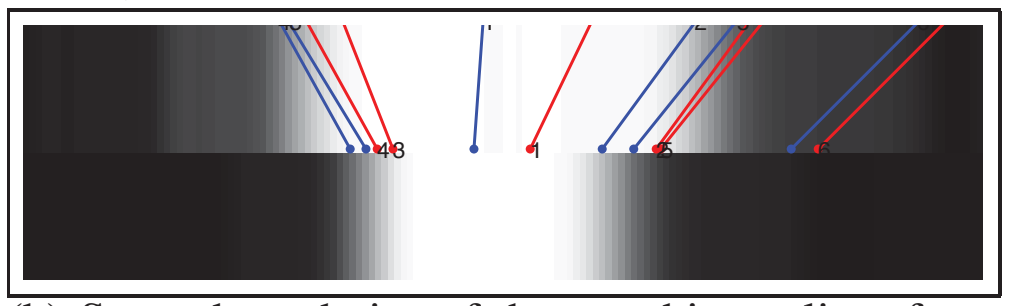

(b) Scope boundaries of the matching salient features after inconsistency elimination

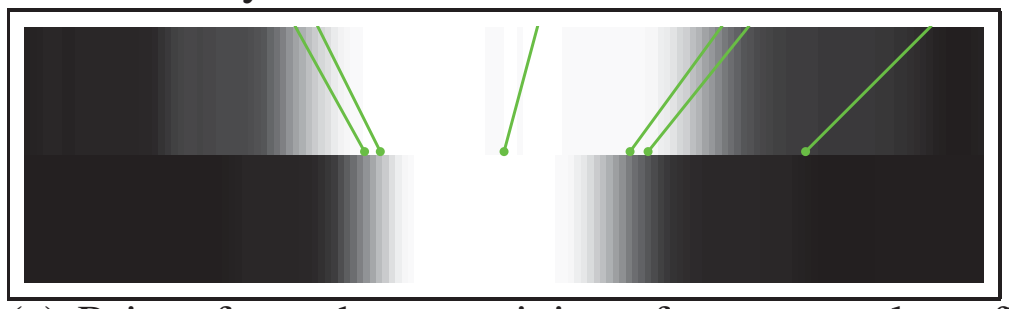

(c) Pairs of matches remaining after temporal conflicts are eliminated – a comparison with (a) will show that some of the pairs have been discarded

**Figure 7: (a) Matching salient feature pairs for the two time series shown in Figure 4, (b) the corresponding scope boundaries (after conflicting scope boundaries have been eliminated), and (c) the pruned subset of matching salient feature pairs**

of the time series, including (a) the scale at which the most discriminating keypoints are located and (b) the size of the temporal range that should be included in the descriptor to help disambiguate the features. In particular, if a given time series contains many similar features, it might be more advantageous to use large descriptors that cover ranges bigger than the scopes of the temporal features: these large descriptors would not only include information that describe the corresponding features, but would also describe the temporal contexts in which these features are located. We investigate the relationship between the high level temporal characteristics of time series and the descriptor length in Section 4.4.

## 3.2 Identification of the Matching Salient Feature Pairs

Once the salient features and their descriptors are identified for a pair of time series, the next step is to identify the matching pairs of salient features across for the two time series being compared.

### *3.2.1 Identification of Dominant Pairs*

This is performed by computing the distance between the feature vectors of each pair of salient points using Euclidean distance and selecting the dominant pairs with small distance:

1. Let $s_{1,i}$ be a salient point in the first time series $X1$, and $s_{2,j}$ be a salient point in the second time series, $X2$. If

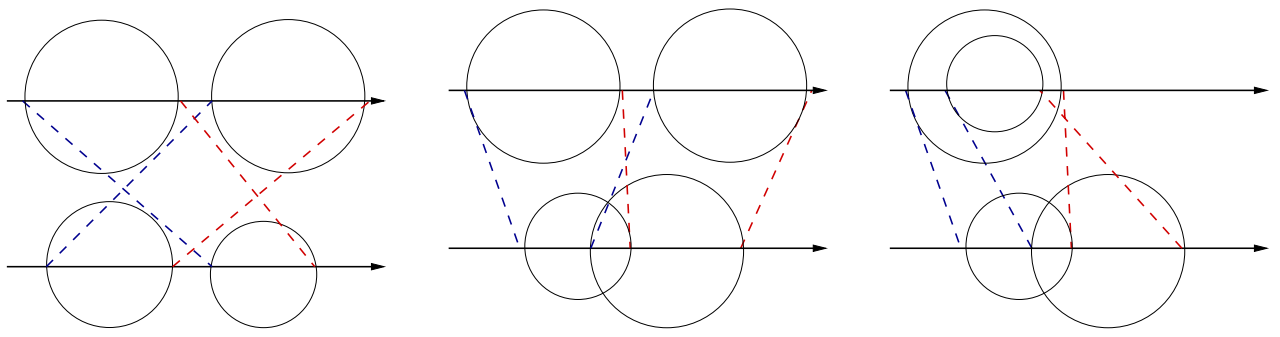

**Figure 8: Example scope boundary conflicts: blue lines mark corresponding starting points of the matching scopes, whereas red lines mark the corresponding end points**

- the amplitude differences between the two salient points is less than a threshold $\tau_a$,
- the ratio of the scales of the two salient points is less than a threshold $\tau_s$, and
- there is no other salient point $s_{2,l}$ satisfying the above two conditions whose descriptor similarity, $sim(s_{1,i}, s_{2,l})$ is also within threshold $\tau_d(> 1)$ of the descriptor similarity, $sim(s_{1,i}, s_{2,j})$; i.e.,

$$sim(s_{1,i}, s_{2,j}) \times \tau_d \leq sim(s_{1,i}, s_{2,l}),$$

then we return the pair $\langle s_{1,i}, s_{2,j} \rangle$ as a matching pair.

Figure 7(a) shows the two time series in Figure 4(a) and the matching pairs of salient features.

### 3.2.2 *Inconsistency Pruning*

Let us take a look at Figure 7(a) again: here we can see that, since the matching algorithm did not impose any constraints on the distances between the time series, the algorithm identified some very distant pairs of matching salient points. Note also that there are many matching pairs that cross each other in time, implying temporal features that are differently ordered in time in two time series. Since, we assume that the transformations on the data has stretched the time differently in the time series, but has not altered the order of the temporal features, we need to eliminate such temporal inconsistencies by pruning matching pairs that conflict with others that are more salient.

Intuitively, we call a set of matchings *consistent* if the corresponding salient features are similarly ordered in both time series. Note that each salient feature has a scope (i.e., starting and end point) defined by the temporal scale of the corresponding salient feature. Therefore, consistent ordering of salient features requires that the scopes of the salient features are similarly ordered. Thus, in order to eliminate temporal inconsistencies, we consider the scope boundaries of the pairs of matching salient features and prune those that imply inconsistent ordering of start and end points of the salient features (see Figure 8). The outline of the process is as follows:

1. For each pair, $\langle f_i, f_j \rangle$ of matching features, we compute

   - an alignment score, $\mu_{align}$, where

   $$\mu_{align}(f_i, f_j) = \frac{(scope(f_i) + scope(f_j))/2}{1 + |center(f_i) - center(f_j)|};$$

   where the scope, $scope(f_i)$ is the temporal length of the feature, $f_i$, and $center(f_i)$ is the position of its center in time – intuitively, we prefer feature pairs consisting of large features that are also position close to each other in time; and

   - a similarity score, $\mu_{sim}(f_i, f_j)$, where

   $$\mu_{sim}(f_i, f_j) = \frac{\mu_{desc}(f_i, f_j)}{\mu_{desc,min}} \times (1 - \Delta_{amp}(f_i, f_j)),$$

   where $\mu_{desc}(f_i, f_j)$ is the matching score between the descriptors of $f_i$ and $f_j$, $\mu_{desc,min}$ is the minimum matching score among all matching pairs, and $\Delta_{amp}(f_i, f_j)$ is the percentage difference between the overall amplitudes of the features within their corresponding scopes – intuitively, we prefer pairs of features that have both similar descriptors and similar average amplitudes.

   Given these, the combined score, $\mu_{comb}(f_i, f_j)$, is computed using F-measure that requires both alignment and similarity scores to be high for a high combined score:

   $$\mu_{comb}(f_i, f_j) = 2 \times \frac{ns_{align}(f_i, f_j) \times ns_{sim}(f_i, f_j)}{ns_{align}(f_i, f_j) + ns_{sim}(f_i, f_j)},$$

   where $ns(f_i, f_j)$ are scores normalized between the range 0 and 1 by dividing each score to the maximum score among all pairs being considered.

2. Next, we consider all pairs of matching features in descending order of their $\mu_{comb}$ scores.
   (a) Let $\langle st_{1,i}, end_{1,i} \rangle$ and $\langle st_{2,j}, end_{2,j} \rangle$ be the scopes of the pair of salient features we are currently considering.
   (b) We *attempt* to insert the time points $st_{1,i}$ and $end_{1,i}$ into a list, $scope_boundary_order_1$ ordered in increasing order of time; similarly we *attempt* to insert the time points $st_{2,j}$ and $end_{2,j}$ into the list, $scope_boundary_order_2$ ordered in increasing order of time.
   (c) Let $rank(st_{1,i})$, $rank(st_{2,j})$, $rank(end_{1,i})$, and $rank(end_{2,j})$ be the corresponding ranks of the time points in their respective time ordered lists.
   (d) If $rank(st_{1,i}) = rank(st_{2,j})$ and $rank(st_{1,i}) = rank(end_{2,j})$, then we confirm the insertion and we keep the pair[1]
   (e) Else, we drop the pair and eliminate these scope boundaries from consideration.

The reason why the feature pairs are considered in descending order of $\mu_{comb}$ scores is that, when a conflict is identified, the new pair –which is relatively less aligned, smaller, and less similar (in terms of descriptors as well as average amplitudes)–can be eliminated without affecting the already committed boundaries.

Figure 7(b) shows the scope boundaries of the maintained matching pairs after the elimination of scope boundary inconsistencies: here blue lines correspond to the starts of the matching scopes and red lines correspond to the ends; note that there are no scope boundaries that cross each other. Figure 7(c) shows the maintained matching pairs after the elimination of inconsistent pairs.

## 3.3 Computation of Locally Relevant DTW Constraints

Once the matching pairs of salient features and their scope boundaries are identified, the next step is to use this information for computing the locally relevant band (width and slope) constraints applicable during the execution of the DTW algorithm.

[1]The process is slightly more complex in that there can be exceptions where the ranks are different, but time values are the same. We also confirm the insertion in these special cases.

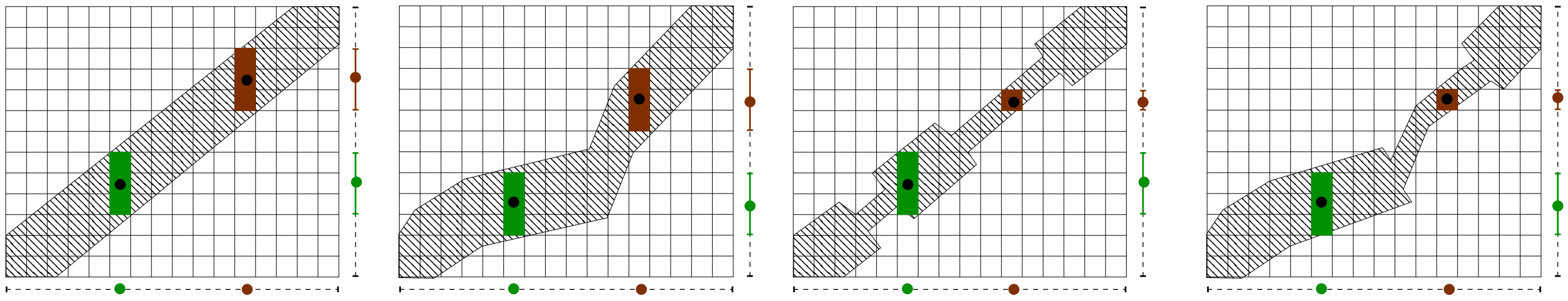

(a) fixed core&fixed width (b) adaptive core&fixed width (c) fixed core&adaptive width (d) adaptive core&adaptive width

**Figure 10: The shape of the bands defined by the (a) *fixed core&fixed width*, i.e., Sakoe-Chiba band, (b) *adaptive core&fixed width*, (c) *fixed core&adaptive width*, and (d) *adaptive core&adaptive width* constraints**

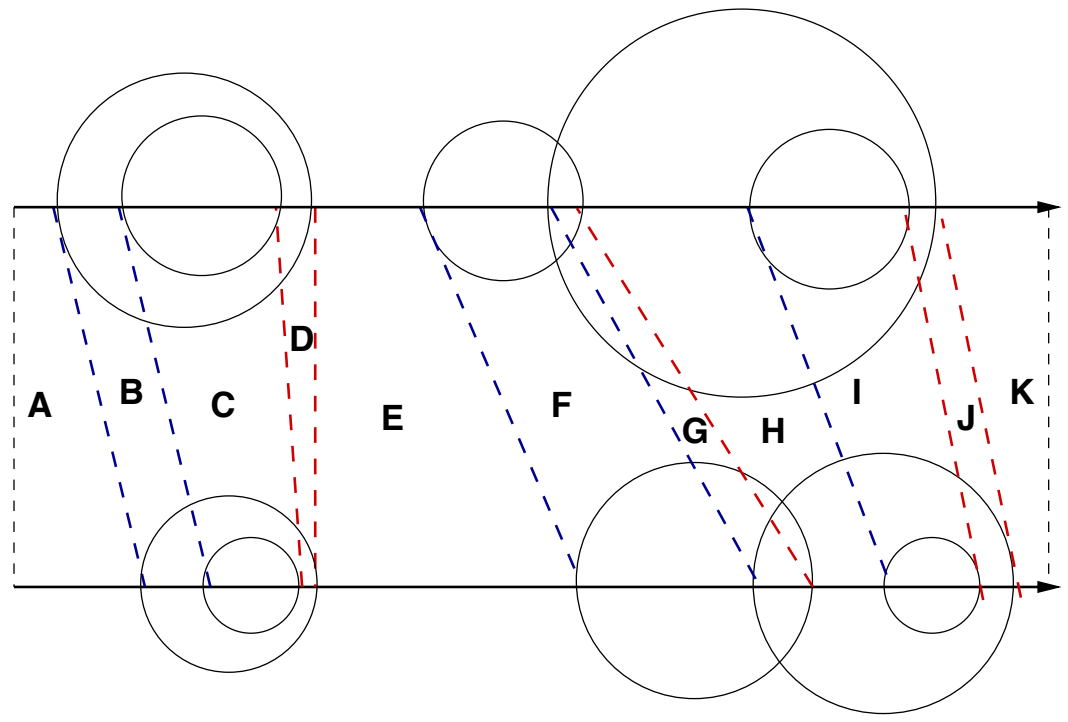

**Figure 9: Scope boundaries**

Consider for example, the two time series in Figure 9, the matching salient features identified on these time series, and the consistently aligned scope boundaries. Note that the resulting scope boundaries partition each time series into a set of consecutive intervals, in this example from interval A to interval K. While the lengths of the corresponding intervals in the two time series differ, it is also clear that each corresponding interval pair in the two time series corresponds to similar regions and thus must have similar characteristics.

### *3.3.1 Adaptive Width Constraints*

Remember from Section 1 and Figure 2(b) that the Sakoe-Chiba band constraint defines a narrow band around the diagonal from which the warp paths can pass. This *fixed core&fixed width approach* is also visualized in Figure 10(a).

The first difficulty with this approach of course is to decide the width of the band. Our first constraint –informed with the salient feature alignments– uses the widths of the resulting intervals to choose a different locally relevant width for each point on the time series (Figures 10(c) and (d)). Let $X$ and $Y$ denote the two time series. For each point $x_i$ in the first time series, a candidate point $y_j$ in the second time series is located and $x_i$ is compared only to the points that are within $\pm\lceil w/2 \rceil$ of $y_j$, where $w$ is the width of the interval in the second time series containing $y_j$.

Note that adaptive width constraints can easily be combined with fixed width constraints (which may be available based on domain knowledge) by imposing lower- and upper-bounds on $w$. A second refinement to the adaptive width constraint formation is to set $w$ to the average of the $r$ intervals around the interval containing $y_j$. This is especially useful in noisy time series where neighboring interval sizes can vary drastically. We will evaluate adaptive width constraints and its refinements in Section 4.

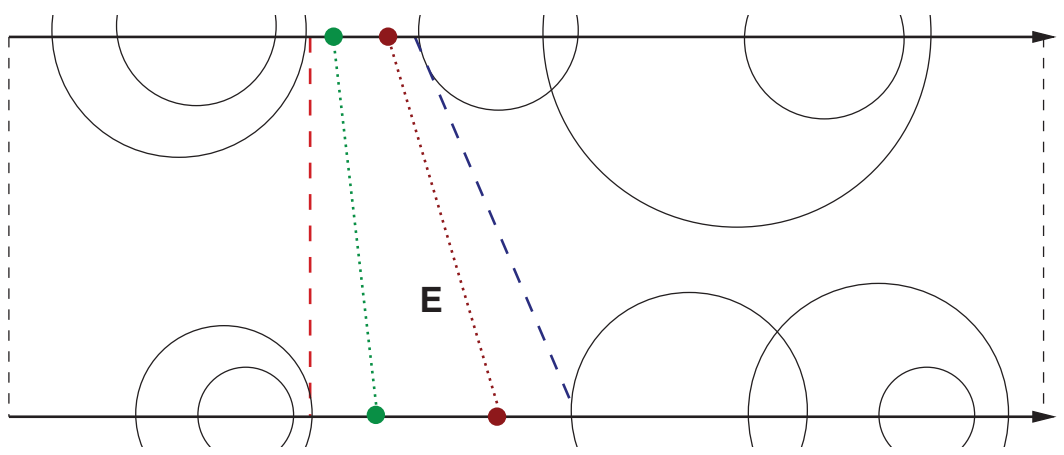

**Figure 11: Sample point alignments within interval E**

### *3.3.2 Adaptive Core Constraints*

If the candidate point, $y_j$, corresponding to $x_i$ is obtained by setting $j = i$, this gives us a fixed diagonal core as shown in Figures 10(a) and (c). Alternatively, we can attempt to locate a candidate point $y_j$ that has a better chance to correspond to $x_i$ using the available structural evidences. Figure 10(b) and (d) shows two bands where the *cores* of the bands are not on the diagonal.

We note that the intervals discovered during the matching and inconsistency removal processes described in the previous section can provide us additional information to help adapt the position of the *core* of this band. Consider Figure 11, which focuses on interval E in Figure 9. As we can see in this figure, we can use the starting and ending points of the corresponding intervals in the two time series to associate to each point in one series a *candidate point* in the other time series. Let $X$ and $Y$ denote the two time series in this example, $st(X,\mathtt{E})$ and $end(X,\mathtt{E})$ denote the start and end points of interval E on the series $X$, and $st(Y,\mathtt{E})$ and $end(Y,\mathtt{E})$ denote the start and end points of interval E on the series $Y$. Let also $x_i$ be a point in series $X$ in interval $E$. The corresponding candidate point, $y_j$, can be computed based on the following equality:

$$\frac{j - st(Y,\mathtt{E})}{end(Y,\mathtt{E}) - st(Y,\mathtt{E})} = \frac{i - st(X,\mathtt{E})}{end(X,\mathtt{E}) - st(X,\mathtt{E})}.$$

Naturally, exceptions occur when one of the intervals is empty. If $end(Y,\mathtt{E}) = st(Y,\mathtt{E})$, this implies that $st(Y,\mathtt{E})$ will be the candidate point for all points in all points in the corresponding interval in $X$. If, on the other hand, $end(X,\mathtt{E}) = st(X,\mathtt{E})$, this may result in a gap in the band. Since such a gap would prevent successful completion of the dynamic programming algorithm, we need to bridge the gap by filling in the missing grid positions.

The resulting *adaptive core&fixed width* and *adaptive core&adaptive width bands* are visualized in Figures 10(b) and (d) respectively: in neither of these cases, the core of the band is aligned with the diagonal; instead, the core follows a path that reflects the alignments implied by the scopes of the salient features of the time series.

| Data Set | Length | # of Series | # of Classes |
|---|---|---|---|
| Gun | 150 | 50 | 2 |
| Trace | 275 | 100 | 4 |
| 50Words | 270 | 450 | 50 |

**Table 1: Data sets used in the experiments**

### *3.3.3 Summary*

The three types of locally relevant constraints described in this section (Figure 10) make different assumptions about the characteristics of the time warp. The *fixed core* constraints assume that the two time series are globally aligned, but locally misaligned. *Adaptive core* constraints, on the other hand, assume that time may be differently skewed at different parts of the two time series and attempt to match this temporal skew based on structural evidences. Note also that since one of the time series, $X$, is used to drive the candidate point search on the other time series, $Y$, the *adaptive width* constraints and the adaptive core constraints result in non-symmetric distance measures. However, the distance can be rendered symmetric by first switching the roles of $X$ and $Y$ and then performing the dynamic programming step using a *combined* band, including grid-cell positions required by both series $X$ and $Y$.

## 3.4 Complexity

As discussed earlier, the computation time for computing the optimal DTW distance involves filling the $N \times M$ DTW grid in $O(NM)$ time and then identifying the optimal warp path in $O(N + M)$ time. In contrast, the computation time of the `sDTW` distance (based on locally relevant constraints) involves the following three components:

- *Time for extracting the salient features:* This component depends on the complexity of the salient feature extraction algorithm.
- *Time for finding matching salient feature pairs and pruning inconsistent pairs:* This is an $O(|S_X| \times |S_Y|)$ step, where $S_X$ is the set of salient features identified in time series $X$ and $S_Y$ is the set of salient features identified in time series $Y$. Note that while this step is also quadratic in time, since $S_X \ll N$ and $S_Y \ll M$, this step is likely to be much faster than time needed to fill the DTW matrix.
- *Time for (partial) filling of the DTW matrix based on the locally relevant constraints and identifying the corresponding warp path:* The worst case time complexity of this step is $O(N \times M)$ (i.e., when no features are identified by the first step to constrain the filling of the DTW grid); however, in practice the cost is likely to be smaller and the time gains will depend on how *tight* the discovered constraints are.

Note that, extraction of salient features is a one-time process. Once these features are extracted, they can be stored and indexed along with the time series and can be re-used repeatedly during various retrieval and classification tasks involving different time series.

# 4. EXPERIMENTS

In this section, we experimentally evaluate the `sDTW` algorithm which relies on *locally relevant constraints* (obtained based on matching salient features) to reduce the DTW computation time.

## 4.1 Data Sets

For assessing the efficiency and effectiveness of `sDTW`, we use time series data sets available at [9]. Table 1 and Figure 12 provide an overview of the time series data sets used for these experiments.

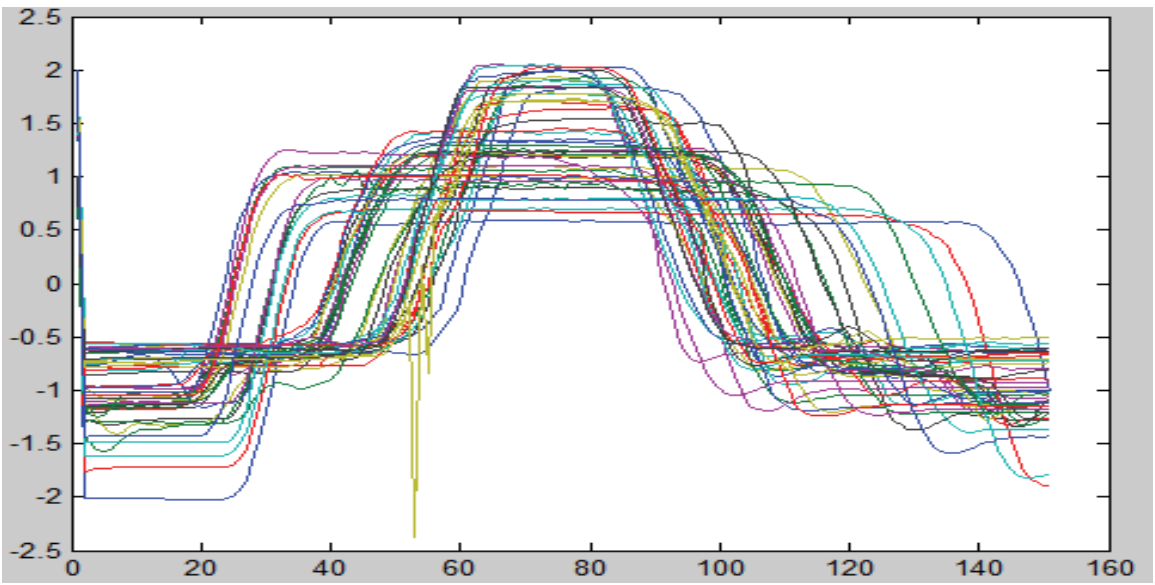

(a) Time series in the `Gun` data set

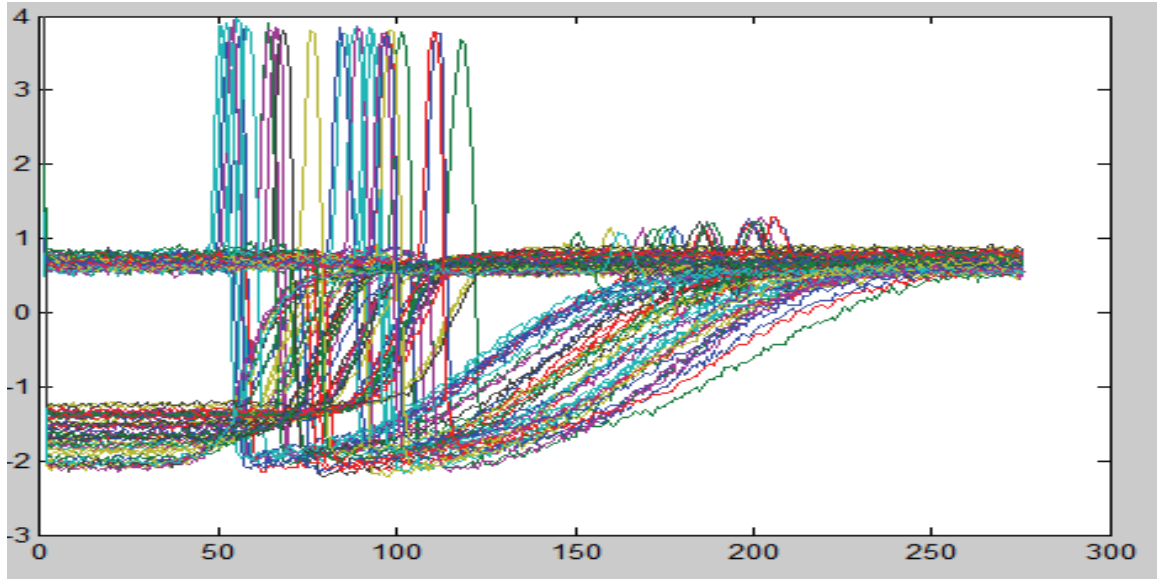

(b) Time series in the `Trace` data set

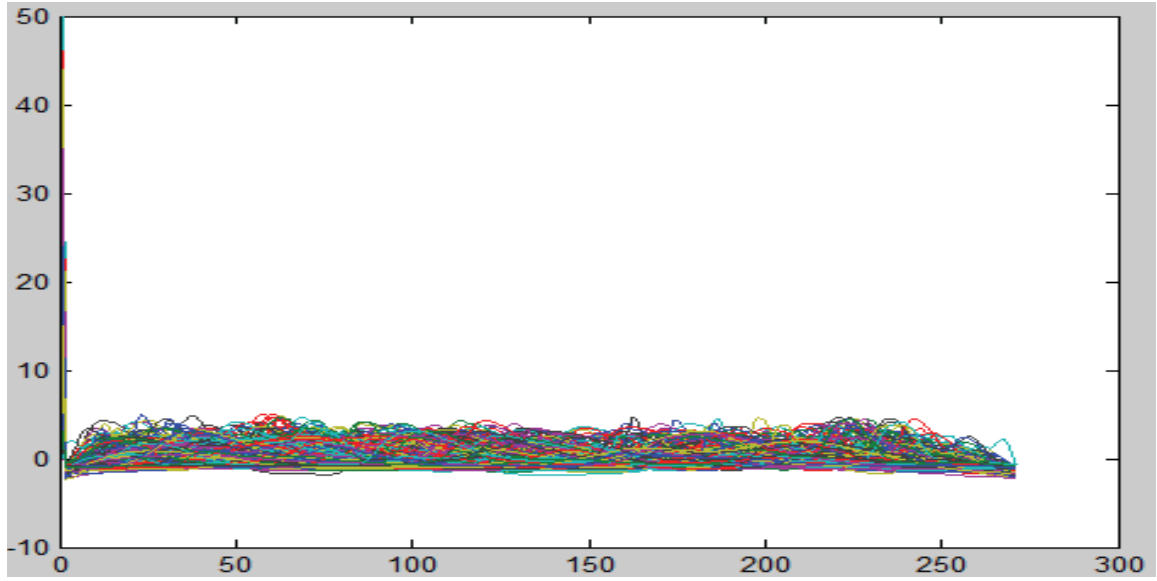

(c) Time series in the `50Words` data set

**Figure 12: Plots of the time series in the three data sets used in the experiments**

| Data Set | Fine | Medium | Rough | Total |
|---|---|---|---|---|
| Gun | 221.2 | 165.4 | 58.9 | 445.5 |
| Trace | 122.1 | 140.0 | 46.6 | 308.7 |
| 50Words | 202.1 | 90.3 | 18.9 | 311.3 |

**Table 2: Average numbers of salient points at three different (fine, medium, and rough) scales in the three data sets**

Before we discuss the evaluation criteria and the results, we also present the salient point distribution in these three data sets in Table 2. In agreement with the plots in Figure 12, the `Gun` data set has the highest number of large scale features; in contrast the series in the `50Words` data set have very few large features.

## 4.2 Evaluation Criteria

In order to assess the effectiveness of various DTW algorithms, we use the following measures:

- *Retrieval accuracy:* The first effectiveness criterion we use is the *top-k* retrieval accuracy; i.e., the overlap between the set of retrieved objects when using the optimal DTW and when using the distance computed using locally relevant constraints:

$$acc_{ret}(k) = \underset{X}{avg} \frac{|top_{DTW}(X,k) \cap top_*(X,k)|}{k}.$$

Here $top_{DTW}(X,k)$ and $top_*(X,k)$ are the top $k$ objects retrieved using the optimal DTW distance and DTW distance constrained with locally relevant constraints, respectively.

- *Distance accuracy:* The second effectiveness criterion we use is the accuracy of the distance estimations relative to the optimal DTW distance (based on the complete DTW grid). In particular we measure

$$err_{dist} = \underset{X,Y}{avg} \frac{\Delta_*(X,Y) - \Delta_{DTW}(X,Y)}{\Delta_{DTW}(X,Y)},$$

where $\Delta_*$ stands for the distance computed based on locally relevant constraints.

- *Classification accuracy:* As shown in Table 1, the set of time series we are using have already been manually labeled into distinct classes. The third effectiveness criterion we use is the $k$ *nearest neighbor* classification accuracy; i.e., the overlap between the nearest neighbor based class labels attached to the objects when using the optimal DTW and when using the distance computed using locally relevant constraints:

$$acc_{cls}(k) = \underset{X}{avg} \frac{|labels_{DTW}(X,k) \cap labels_*(X,k)|}{|labels_{DTW}(X,k) \cup labels_*(X,k)|}.$$

Here $labels_{DTW}(X,k)$ and $labels_*(X,k)$ are the sets of class labels attached to an object $X$ by $k$ nearest neighbor algorithm, based on the optimal DTW distance and DTW distance constrained with locally relevant constraints, respectively. Note that the $k$ nearest neighbor algorithm can attach more than one label to the time series, $X$, if there are more than one class labels with the same maximum count in the result set, $top_-(X,k)$.

As discussed earlier, the computation time for optimal DTW distance involves filling the DTW grid and identifying the optimal warp path. The computation time of the `sDTW` distance based on locally relevant constraints, on the other hand, involves time (a) for the extraction of the salient features, (b) for finding matching salient feature pairs and pruning inconsistent pairs, and (c) for (partially) filling the DTW matrix based on the locally relevant constraints and identifying the corresponding warp path. As also described earlier, the first task (a) is performed **only once** per time series, whereas tasks (b) and (c) need to be repeated for each pair of time series that are compared. Also, for these data sets, the salient point extraction was negligible ($\sim$ 0.7ms for `gun`, $\sim$ 1ms for `trace`, and $\sim$ 3ms for `50Words` data sets).

Therefore in this section, we focus on the average execution times for tasks (b) and (c) and report time gain as

$$timegain = \frac{time_{DTW} - time_*}{time_{DTW}},$$

where $time_*$ denotes the time to execute tasks (b) and (c) for DTW distance computation with locally relevant constraints.

## 4.3 Approaches

In this section, we consider the following four approaches:

- *Full DTW (dtw):* This is the optimal DTW which uses the full grid.
- *Fixed core&fixed width (fc,fw):* This is the Sakoe DTW which limits the search to a fixed diagonal band, where each point in the first time series is compared only to $w\%$ of the points in the second time series. We consider three different values of $w$: $6\%$, $10\%$, and $20\%$.

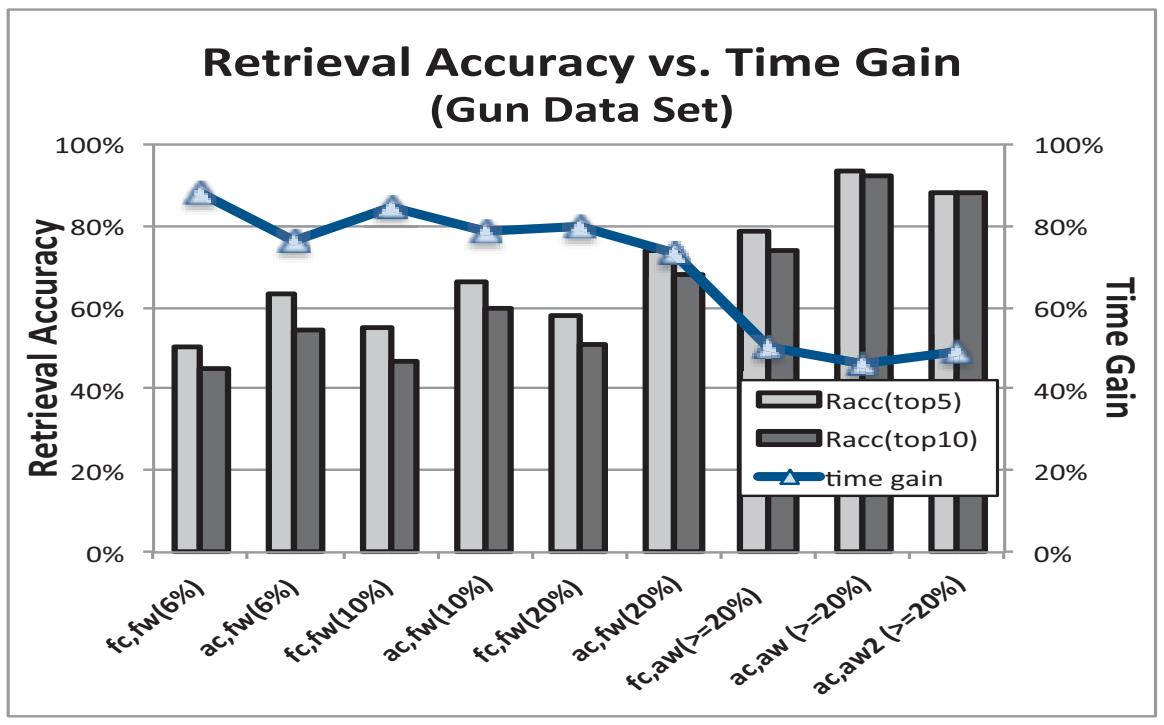

(a) Gun data set

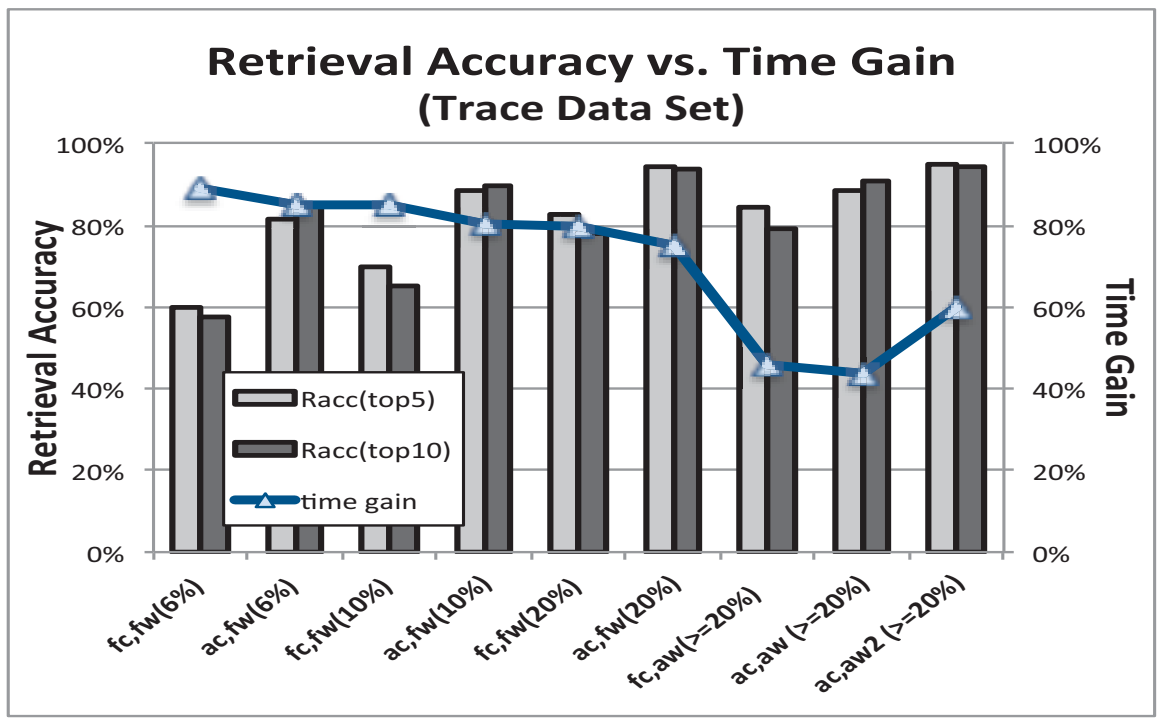

(b) Trace data set

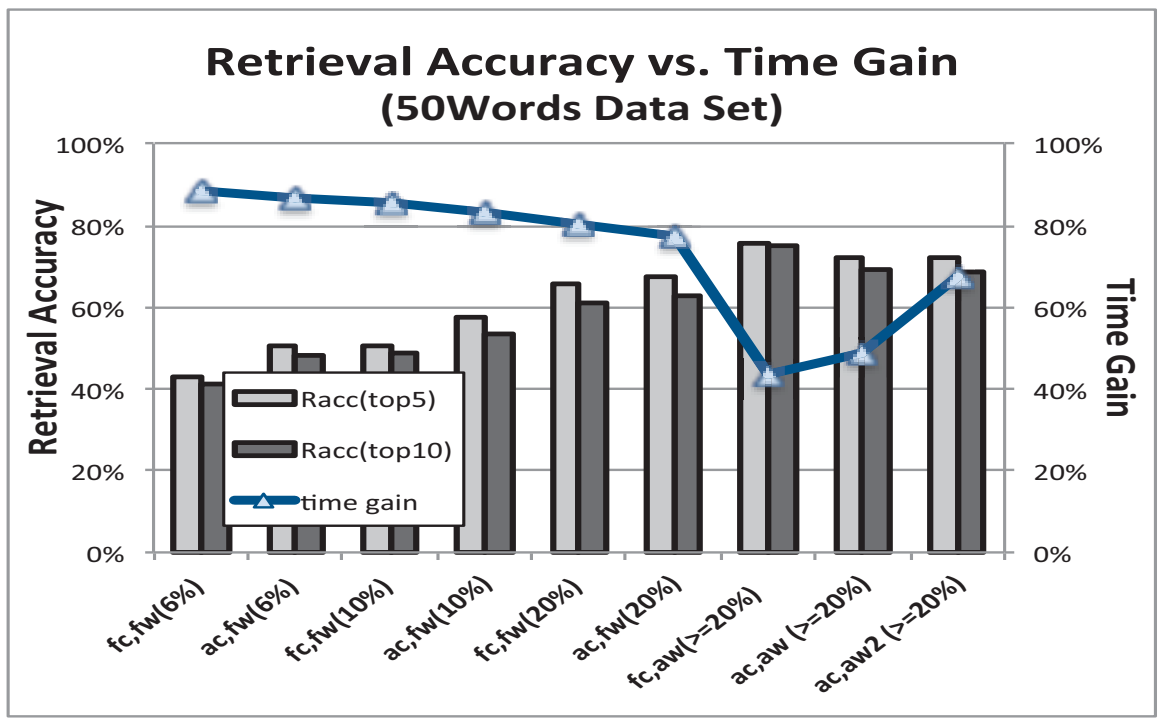

(c) 50Words data set

**Figure 13: Retrieval accuracy and time gain results for different data sets and for different values of $k$**

- *Fixed core&adaptive width (fc,aw):* In this case, the width is not fixed, but adapts (with a lower-bound of 20%) based on local features of the second time series, as described in Section 2.1.3.
- *Adaptive core&fixed width (ac,fw):* This is similar to Sakoe DTW, but the core is not necessarily diagonal; instead it is adapted as described in the paper. Here, we also consider three different values of $w$: $6\%$, $10\%$, and $20\%$.
- *Adaptive core&adaptive width (ac,aw and ac2,aw):* In this case, both the core and the width adapt. We experiment with two versions: the first version uses the size of the local interval to determine the width, $w$; whereas, the second version averages the sizes of the previous, current, and next intervals.

Experiments reported in this section were collected on an Intel Core 2, Quad CPU 3GHz machine, with 8GB RAM, running Ubuntu 9.10 (64bit). All code was implemented using Matlab. For the baseline *fixed core&fixed width* schemes, we used the DTW code available at [9]. For 1D salient point detection, we modified the SIFT code obtained from [19]. Experiments were executed using Matlab 7.8.0. Unless specified otherwise, we use feature descriptors with 64 bins, obtained using $o = (\lfloor log_2(N) \rfloor - 6)$ octaves, where $N$ is the length of the time series, each with $s = 2$ levels. Also, in these experiments the value of $\epsilon$ used for salient feature search was set to $0.96\%$ (see Section 3.1.2).

## 4.4 Results and Discussions

**Top-$k$ Retrieval Accuracy vs. Time Gain.** Figure 13 presents the top-5 and top-10 retrieval accuracies of various algorithms as well as the corresponding time gains for different data sets. As can be seen here, the relative behavior of the various algorithms are similar in all three data sets and for both targets $k$: (a) as expected, for *fixed core&fixed width* (*fc,fw*) algorithms, the larger is the value of $w$, the more accurate the results are; (b) on the other hand, significant gains in accuracy is obtained when the core is adapted (*ac,fw*) and the accuracy is further boosted when also adapting the width (*ac,aw*). The accuracy gains are especially pronounced for the larger `50Words` data set, where finding top-5 or top-10 results is more difficult. The figure also shows that, while fixed width and fixed core algorithms tend be slightly faster than their adaptive counter-parts, the gains relative to full DTW are still high. In particular, the *adaptive core&adaptive width* algorithm with neighbor averaging (*ac, aw 2*) provides high accuracy with large time gains.

**DTW Distance Error vs. Time Gain.** Figure 14 confirms the above results by focusing on the errors in the DTW estimates provided by the various algorithms for each data set. As can be seen in the figure, the *fixed width&fixed core* algorithms tend to result in extremely high errors (especially in the `gun` data set, which has only two classes of time series). While the ranges of errors are different for different data sets, the relative behaviors of the algorithms are similar – with the notable exception of *fixed core&adaptive width* algorithm (*fc, aw*) whose relative performance depends on the degree of major shifts and skews in the data and which provides the smallest error for the `50Words` data set which does not contain major shifts, but only minor deformations around the diagonal.

Note that since time series in a given class are more likely to be similar to each other then the series in the overall data set, achieving high accuracy within the same class is likely to be more difficult. Therefore, in Figure 15, we present experiment results reporting intra-class distances for the `trace` data set, which has 4 classes, each with $\sim 25$ series. As can be seen here, indeed, fixed core algorithms are especially error prone when provided a data set with very similar objects (with distance errors up to $1000\%$), whereas adaptive core algorithms bring errors down to $\sim 10\%$ range.

**Classification Accuracy vs. Time Gain.** Figure 16 re-confirms the above results by focusing on the classification accuracy. Since it has the largest number of objects (450) and classes (50), here we focus on the `50Words` data set (the $k$ nearest neighbor classification accuracies are higher for all other algorithms on the other two data sets). As can be seen here, similarly to the top-$k$ retrieval and DTW distance accuracy results, adaptive core, and adaptive width algorithms improve the classification accuracy.

**Execution Time Analysis.** Figure 17 shows the contribution of matching (including inconsistency elimination) and dynamic programming steps to the overall DTW computation time. The *fixed width&fixed core* algorithms do not have any matching computation overhead. In contrast, adaptive algorithms need to first identify matches and the complexity of this step depends on the number of features in the time series. As can be seen here, however, matching is a small proportion of the overall work and time is spent mostly

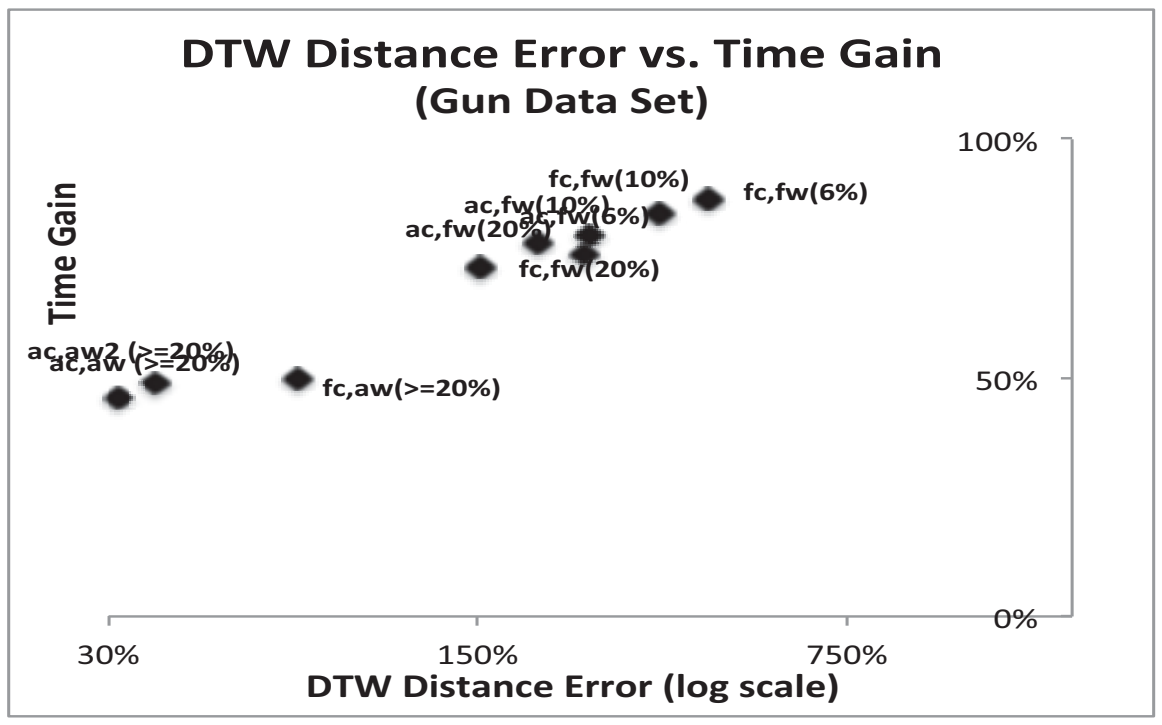

(a) Gun data set

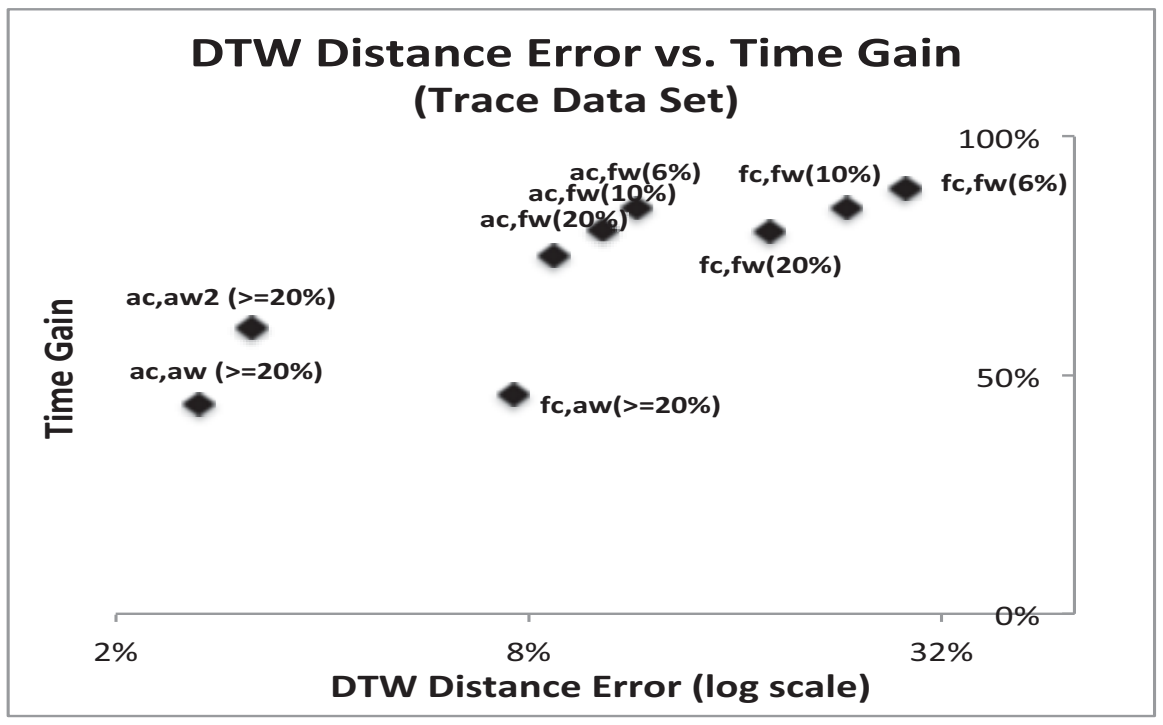

(b) Trace data set

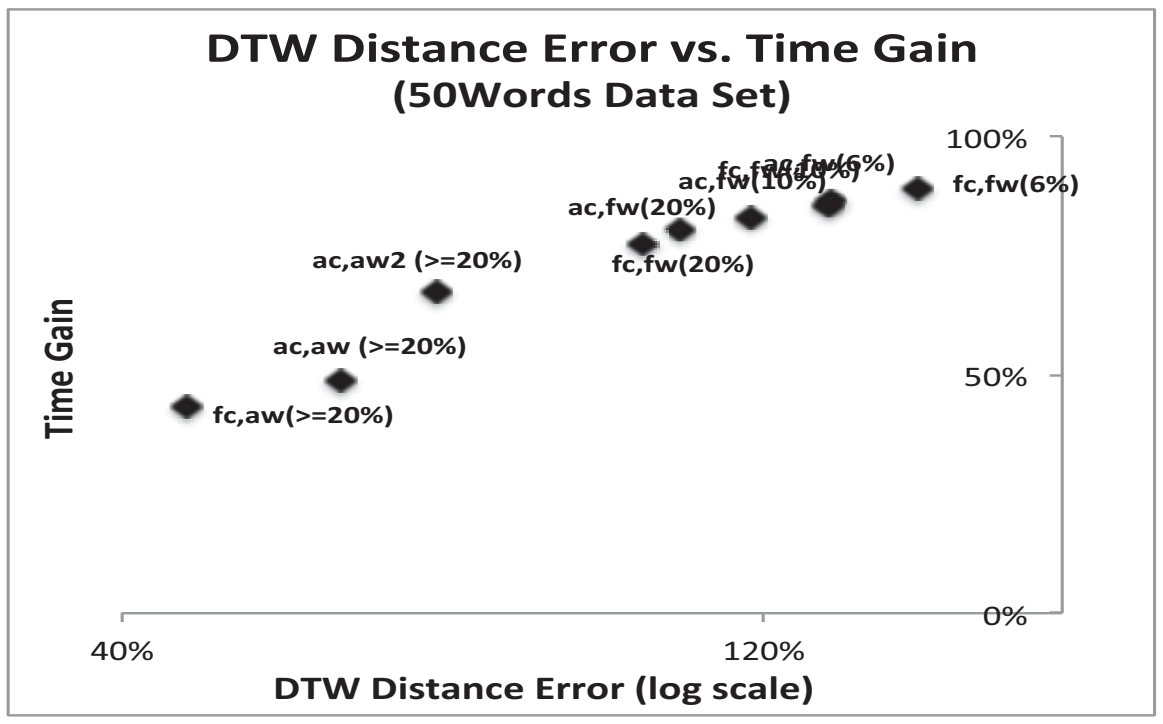

(c) 50Words data set

**Figure 14: Distance error vs. time gain results for different data sets (distance error results are presented on the $x$ axis, whereas time gain results are presented on the $y$ axis)**

during the dynamic programming step (the shares of matching and inconsistency removal were even lower for the other data sets).

**Impact of the Descriptor Length.** In all the experiments reported so far, we have used feature descriptors with 64 bins. In Figure 18, we analyze the impact of the descriptor length (varied between 4 and 128) on distance estimations, top-10 accuracy, and speedup in more detail. Before we present the results, it is important to remember the following relationships between descriptor lengths and temporal features discussed in Section 3.1.2:

- The same descriptor size would cover different temporal ranges in different time scales. Salient points corresponding to small temporal features are located at scales close to the original time scale, whereas large features are often located in scale spaces corresponding to reduced series. Hence, the size of the descriptor needed to represent a temporal feature is often independent of the size of the feature.
- On the other hand, the more discriminating features a time

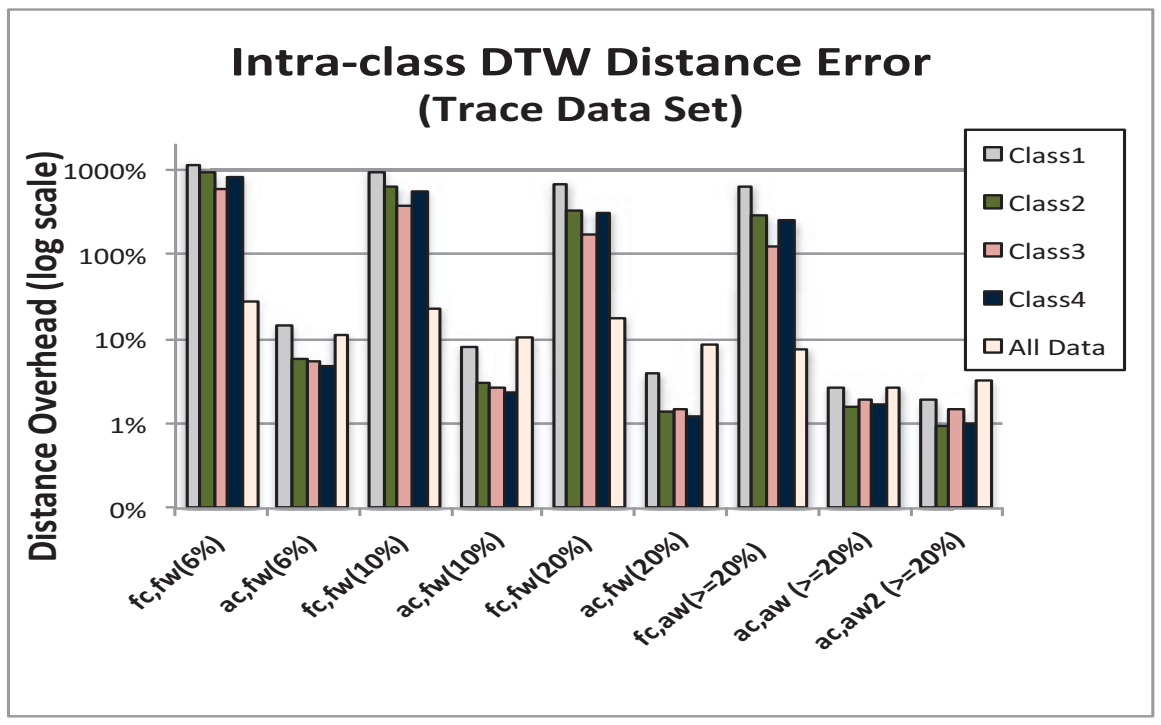

**Figure 15: Intra-class distance errors for the `Trace` data set (this data set has 4 classes each with $\sim 25$ time series)**

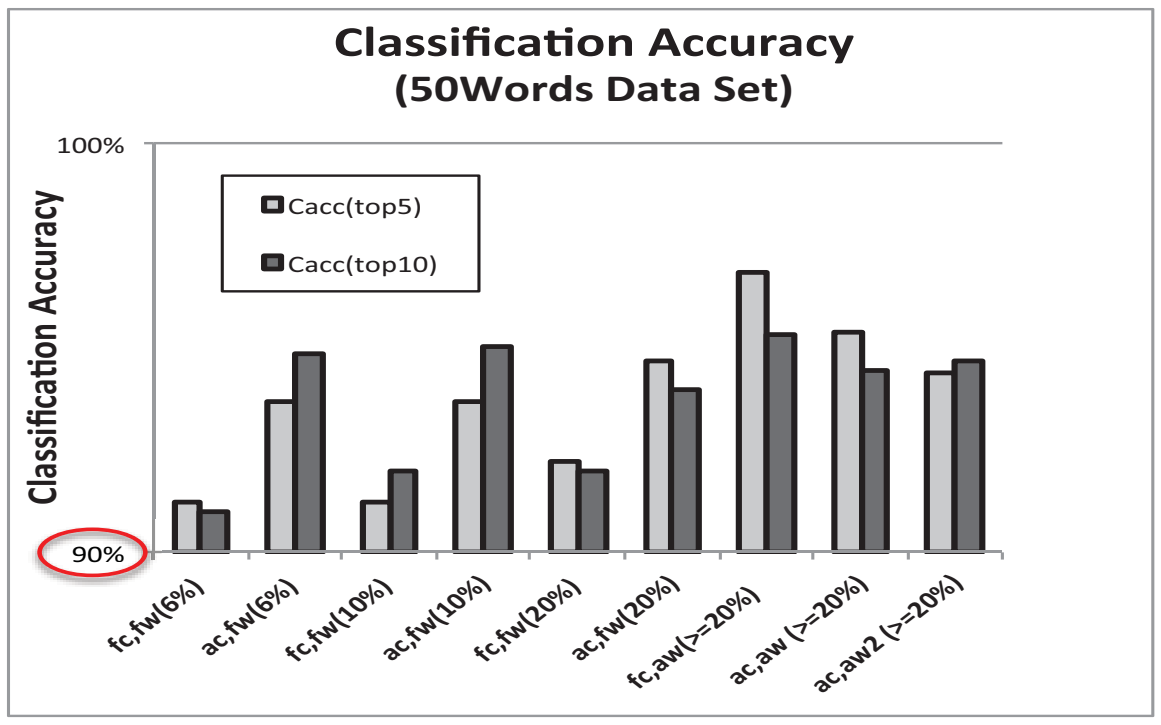

**Figure 16: Top-5 and top-10 classification accuracy (vs. time gain) results for the `50Words` data set (which has 50 classes)**

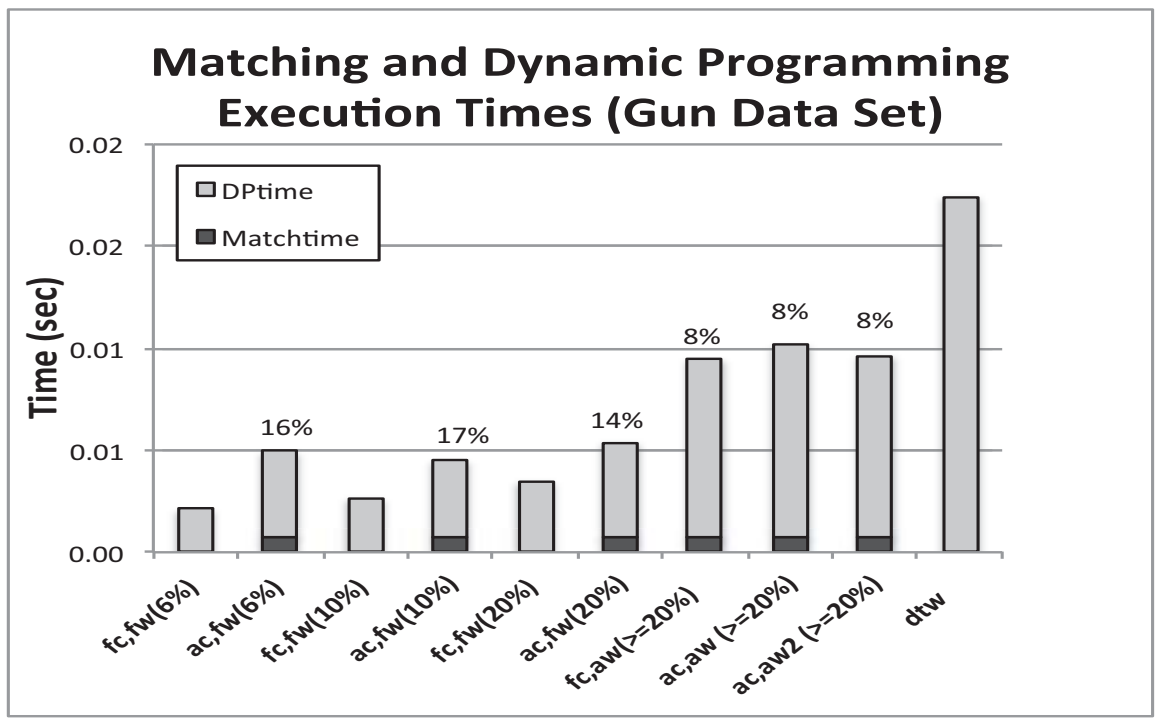

**Figure 17: Execution times for matching/inconsistency removal and dynamic programming (time spent for matching are also highlighted for the adaptive algorithms)**

series has, the smaller descriptors it needs. In time series where the features are not sufficiently discriminating, we may need descriptors extracted from an area larger than the feature itself to help provide temporal context *around* the feature that can help improve feature alignments.

Given these, we can interpret the results presented in Figure 18. While the general trends across the three data sets are similar, there are a few key differences that reflect the characteristics of the underlying time series:

- *Adaptive core&fixed width (ac,fw):* For all three data sets, *adaptive core&fixed width* algorithms do not function as well with very small descriptors. In `Gun` and `Trace` data sets the best descriptor length is around 32; larger descriptors do not help. In the `50Words` data set, however, larger descriptors provide better accuracy. The key reason for this difference is that, as we have seen in Figure 12 and Table 2, `50Words` data set does not have large, discriminating temporal features which can be differentiated from each other using small number of descriptor bins. The lack of highly discriminating features in the `50Words` data set implies that we need descriptors covering an area larger than the features themselves to disambiguate them from each other.
- *Fixed core&adaptive width (fc,aw):* In all three data sets, when using the *fixed core&adaptive width* (fc,aw) approach, the smallest descriptor length (i.e., 4) provides the best accuracy. This gain, however, comes with a drop in the time gains. Significant time gains and good accuracy can be obtained by using large descriptors that provide temporal context to the discovered features.
- *Adaptive core&adaptive width (ac,aw and ac2,aw):* As we have also observed before, the best accuracy/speed-up tradeoffs are obtained when using *adaptive core&adaptive width* techniques. For the `Gun` data set, very small descriptors are again ineffective. In terms of top-10 accuracy, descriptor size 4 provides very good results in `Trace` and `50Words` data sets, but this comes with a drop in time gains. As before, larger descriptors provide better accuracy and speedup by providing temporal context to the features on the time series.

We can summarize these as follows: The appropriate descriptor length depends on the discrimination power of the features in the data set. The more discriminating the features are, the smaller number of descriptor bins we need. While they tend to provide the best overall accuracy, the adaptive core algorithms rely on proper alignment of the temporal features in the time series and thus are relatively more sensitive to variations in descriptor lengths.

# 5. CONCLUSIONS

In this paper, we recognize that the time series that are being compared often carry structural evidences that can help identify *locally relevant* constraints that can prune unnecessary work during dynamic time warp (DTW) distance computation. The proposed `sDTW` approach first identifies robust salient features in the given pair of time series using a scale invariant transform and then seeks consistent alignments of these salient features across the time series. These salient alignments provide evidences regarding how to constraint the search for optimal warp path. We have proposed three different constraint types based on different assumptions on the structural variations in the data set and experimentally evaluated these over pre-classified data sets. Experiment results have shown that the proposed locally-relevant constraints based on salient features help improve accuracy in DTW distance estimations.

# 6. ACKNOWLEDGMENTS

Authors thank Prof. Eamonn Keogh who shared with them time series datasets and the code for *fixed core&fixed width* DTW.

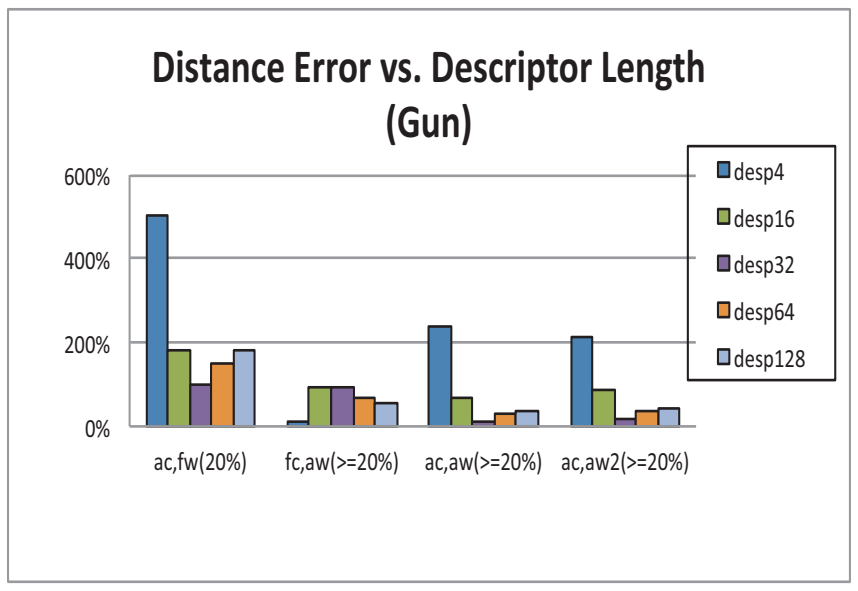

(a) Distance error (Gun)

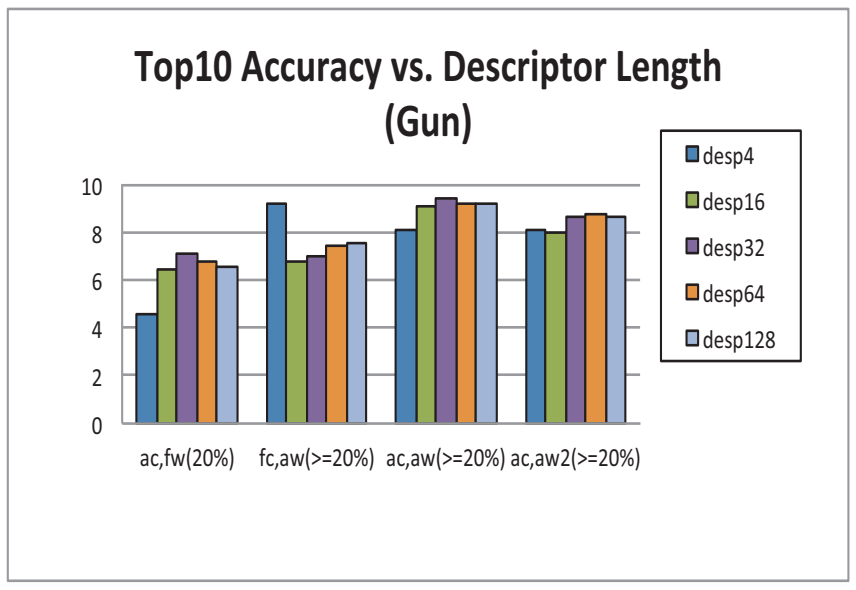

(b) Top-10 accuracy (Gun)

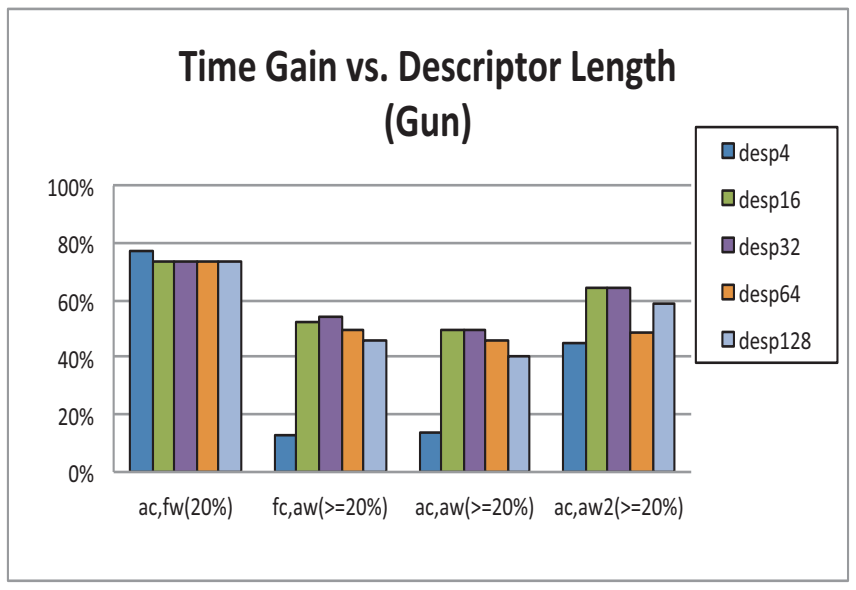

(c) Time gain (Gun)

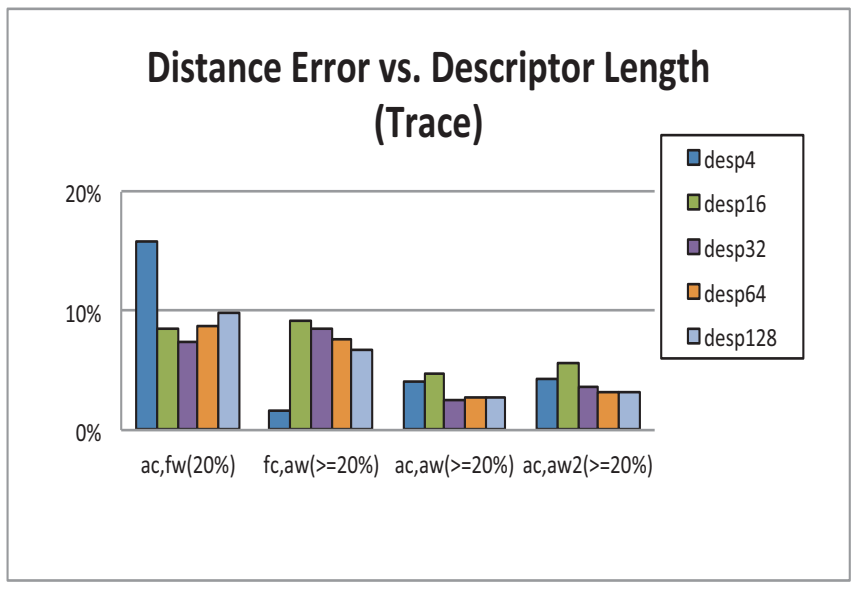

(a) Distance error (Trace)

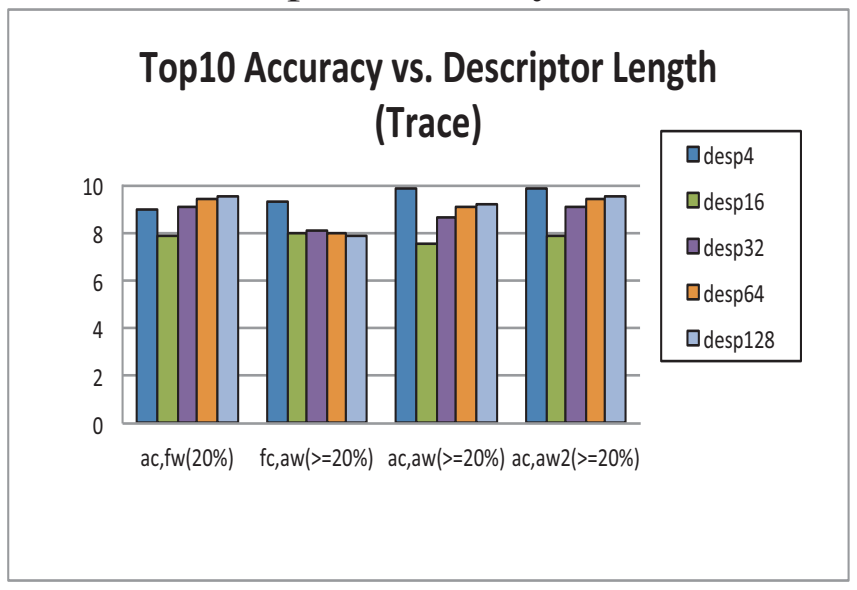

(b) Top-10 accuracy (Trace)

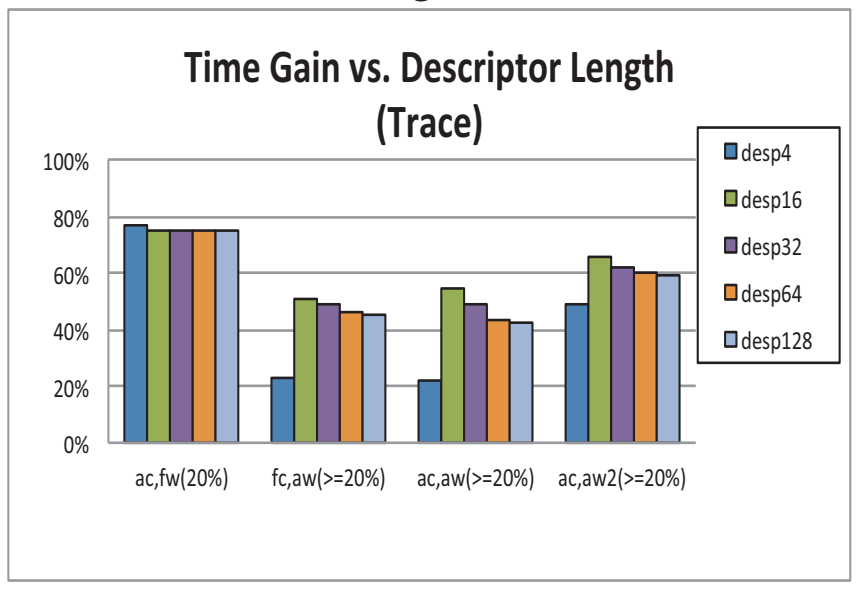

(c) Time gain (Trace)

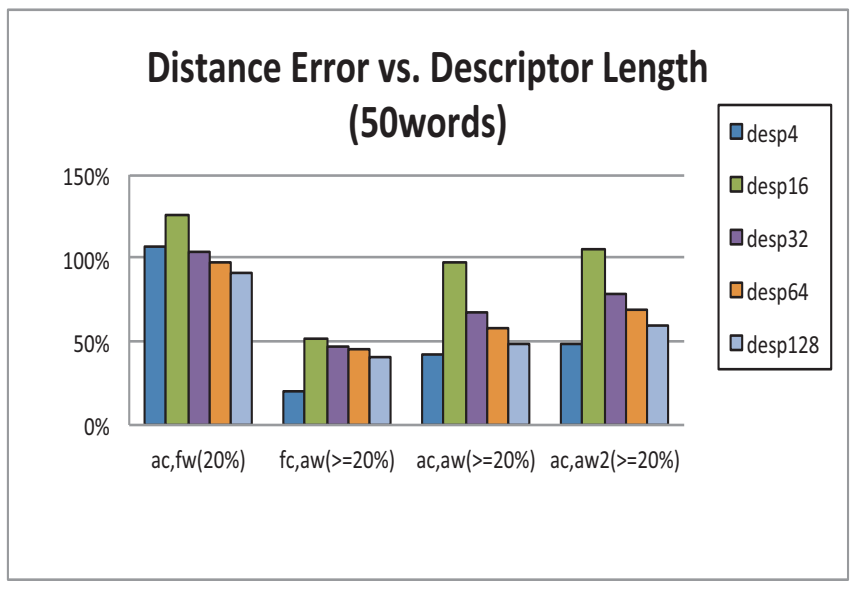

(a) Distance error (50Words)

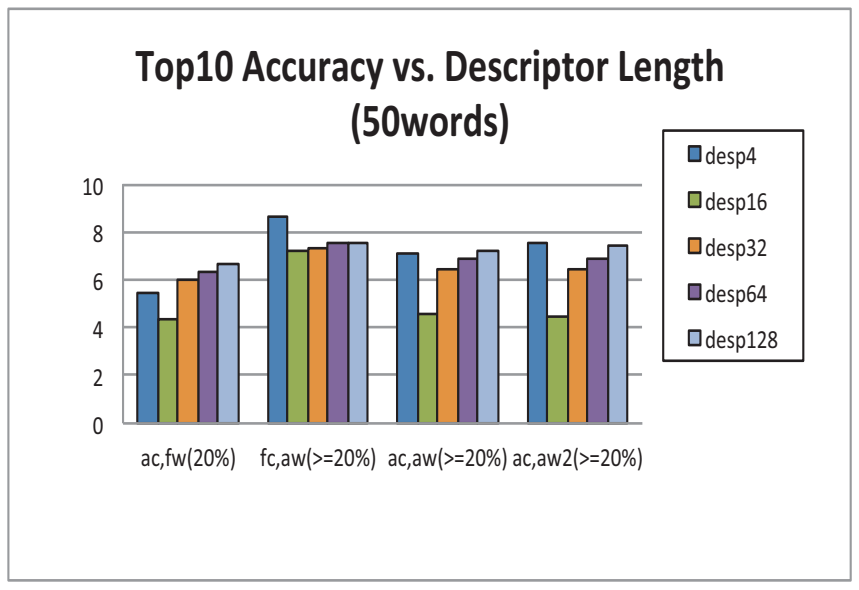

(b) Top-10 accuracy (50Words)

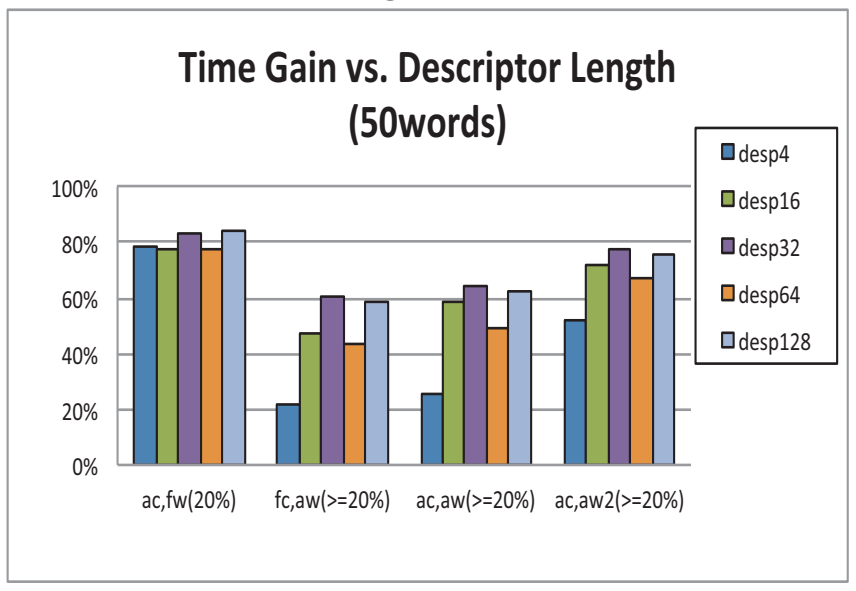

(c) Time gain (50Words)

**Figure 18: Impact of the descriptor length on distance estimations, top-10 accuracy, and speedup**